\documentclass[
reprint,
superscriptaddress,
showpacs,
preprintnumbers,
nofootinbib,
nobibnotes,
bibnotes,
amsmath,
amssymb, 
aps,
prd,
]{revtex4-2}

\usepackage[utf8]{inputenc}
\usepackage{soul}
\usepackage{graphicx}  
\usepackage{dcolumn}   
\usepackage{bm,relsize}        
\usepackage{amssymb,amsmath,mathrsfs}
\usepackage{textcomp}
\usepackage{wasysym}
\usepackage{slashed}
\usepackage{multirow}
\usepackage{lipsum, color}
\usepackage[colorlinks=true,allcolors=purple]{hyperref}
\usepackage[usenames,dvipsnames,svgnames]{xcolor}
\usepackage{booktabs}
\usepackage{xfrac}
\usepackage[capitalize]{cleveref}
\usepackage{isotope}
\usepackage{accents}
\usepackage{titlesec}
\usepackage{colortbl}
\usepackage{pifont}
\usepackage[normalem]{ulem}
\usepackage{braket}
\titlespacing{\subsection}{0pt}{0.1in}{0.1in}

\definecolor{wildstrawberry}{rgb}{1.0, 0.26, 0.64}

\definecolor{forestgreen}{HTML}{228B22}

\newcommand{\lepton}{{\ell}}

\newcommand{\be}{\begin{equation}}
\newcommand{\ee}{\end{equation}}
\def\ltap{\ \raise.3ex\hbox{$<$\kern-.75em\lower1ex\hbox{$\sim$}}\ }
\def\gtap{\ \raise.3ex\hbox{$>$\kern-.75em\lower1ex\hbox{$\sim$}}\ }
\def\lsim{\ \raise.3ex\hbox{$<$\kern-.75em\lower1ex\hbox{$\sim$}}\ }
\def\gsim{\ \raise.3ex\hbox{$>$\kern-.75em\lower1ex\hbox{$\sim$}}\ }
\renewcommand\Re{\operatorname{Re}}

\newcommand{\el}[1]{\textcolor{Green}{#1}}     
\newcommand{\muon}[1]{\textcolor{Purple}{#1}} 
\newcommand{\had}[1]{\textcolor{Gray}{#1}}  

\newcommand{\beq}{\begin{equation}}
\newcommand{\eeq}{\end{equation}}
\newcommand\diag{\operatorname{diag}}
\newcommand\cC{\mathcal{C}}
\newcommand\cS{\mathcal{S}}

\newcounter{modelnum}

\begin{document}

\title{New physics in multi-lepton tau decays}

\author{Yohei Ema}
\affiliation{Institute for Fundamental Theory, Physics Department, University of Florida, Gainesville, FL 32611, USA}

\author{Patrick J.~Fox}
\affiliation{Theory Division, Fermilab, Batavia, IL 60510, USA}

\author{Matheus Hostert}
\affiliation{Department of Physics and Astronomy, University of Iowa, Iowa City, IA 52242, USA}

\author{Tony Menzo}
\affiliation{Theory Division, Fermilab, Batavia, IL 60510, USA}
\affiliation{Department of Physics and Astronomy, University of Alabama, Tuscaloosa, AL 35487, USA}

\author{Maxim Pospelov}
\affiliation{School of Physics and Astronomy, University of Minnesota, Minneapolis, MN 55455, USA}
\affiliation{William I. Fine Theoretical Physics Institute, School of Physics and Astronomy, University of Minnesota, Minneapolis, MN 55455, USA}

\author{Anupam Ray}
\affiliation{The Arthur B. McDonald Canadian Astroparticle Physics Research Institute, Department of Physics, Engineering Physics, and Astronomy, Queen’s University, Kingston, ON, K7L3N6, Canada}
\affiliation{Perimeter Institute for Theoretical Physics, Waterloo, ON, N2L2Y5, Canada}

\author{Jure Zupan}
\affiliation{Department of Physics, University of Cincinnati, Cincinnati, Ohio 45221, USA}

\date{\today}
\begin{abstract} 
Dark particles with lepton-flavor-violating couplings to the tau lepton can induce rare neutrinoless $\tau$ decays with large final state multiplicities. We study models where transitions of the type $\tau^\pm\to \ell^\pm\,\phi$, with $\phi$ a light new particle, initiate a chain of decays in the dark sector that terminate with decays into electrons, muons, or pions.
These decay cascades appear as rare five or even seven-body $\tau$ decays with multiple reconstructable resonances. We survey several representative models: kinetically mixed dark photon, gauged $L_i-L_j$ models, and other more exotic charge assignments such as chiral $U(1)'$ extensions of the Standard Model. The main new ingredient is the possibility of flavor violation at very high scales. In these models, a number of channels that have not yet been searched for experimentally, such as $\tau\to 5\mu$, $\tau\to 3\mu\,2e$, $\tau\to \mu\,4e$, and hadronic channels like $\tau\to \mu\,4\pi$, typically dominate over  the previously-considered signatures such as $\tau \to 3\mu$. While some of the models, such as the gauged $L_i-L_j$ ones, also contain more challenging channels with missing energy due to decays to neutrinos, they can still be searched for via fully visible channels.
\end{abstract}

\maketitle
\tableofcontents
\section{Introduction} \label{sec:intro}

Searches for light new physics, commonly referred to as the searches for ``dark sectors'', now constitute a significant part of the experimental effort in particle physics, both at colliders, beam dump facilities, and at flavor factories (see, {\em e.g.},~\cite{Beacham:2019nyx,Battaglieri:2017aum,Lanfranchi:2020crw,Cirelli:2024ssz} and references therein). 
Among possible new physics probes, rare decays of mesons and charged leptons play an especially important role since their SM decay widths are suppressed by the $W$ mass, $\Gamma \propto 1/m_W^4$. 
The sensitivity to light new physics is therefore parametrically enhanced (see, e.g., \cite{Goudzovski:2022vbt,Tammaro:2025zso,MartinCamalich:2025srw}), as long as the dark sector particles are light enough to be produced either in rare muon \cite{Hostert:2023gpk,Greljo:2025ljr,Hill:2023dym,Ema:2022afm,Ema:2022afm,Knapen:2024fvh,Fox:2024kda,Knapen:2023iwg,Echenard:2014lma,Perrevoort:2018ttp,Panci:2022wlc,Escribano:2020wua,Calibbi:2020jvd,Bigaran:2025uzn,Calibbi:2025rxn,Bauer:2021mvw,Smolkovic:2019jow} or tau decays \cite{Bigaran:2025uzn,Ema:2025bww,Calibbi:2025rxn,Bauer:2021mvw,Escribano:2020wua,Calibbi:2020jvd,Smolkovic:2019jow,Heeck:2025jfs}, in rare kaon decays \cite{Ziegler:2020ize,Hostert:2020gou,Hostert:2020xku,Goudzovski:2022vbt,Delaunay:2025lhl,Calibbi:2025rxn,Datta:2020auq,Gori:2020xvq,Calibbi:2025rxn,Bauer:2021mvw,Smolkovic:2019jow,Calibbi:2016hwq,Ema:2016ops,Alda:2025uwo,MartinCamalich:2020dfe}, rare decays of $B$ and $D$ mesons \cite{Calibbi:2016hwq,Ema:2016ops,MartinCamalich:2020dfe,Calibbi:2025rxn,Alda:2025uwo,Altmannshofer:2023hkn,Bauer:2021mvw,Smolkovic:2019jow} or hyperons~\cite{Alonso-Alvarez:2021oaj}.  

Experimentally, there are significant differences between {\em i)} rare decays in which dark sector particles are stable on detector time scales, and thus appear in a detector as a missing energy signature, and {\em ii)} new physics models in which dark sector particles decay back to visible particles within detector volume. 
An example of the first type of a decay is $\tau \to \mu a$, where $a$ is a QCD axion, stable on cosmological time scales \cite{Calibbi:2020jvd}.
The current world best limit for this decay is  $\mathcal{B}(\tau \to \mu a)< 6 \times 10^{-4}$~\cite{Belle-II:2022heu} (see also \cite{Bryman:2021ilc}). 
This should be contrasted with exotic neutrinoless tau decays that result in a fully visible final state.
These decays are smoking-gun signatures of lepton-flavor violation (LFV) and therefore of physics beyond the Standard Model (SM).
For example, decays such as $\tau \to 3\mu$ and $\tau\to e e\mu$ are currently some of the most well constrained branching ratios of the tau lepton at $\mathcal{B}(\tau \to 3\mu)<1.5$\,--\,$1.9\times 10^{-8}$~\cite{Hayasaka:2010np,BaBar:2010axs,LHCb:2014kws,ATLAS:2016jts,CMS:2020kwy,Belle-II:2024sce}.
Such decays can arise in a large class of BSM scenarios, for instance through on-shell mediators as studied in, e.g., Refs.~\cite{Heeck:2016xkh,Farzan:2015hkd,Heeck:2017xmg,Cheung:2021mol,Cui:2021dkr}, or through a variety of higher-dimensional operators~\cite{Langacker:2000ju,Chiang:2011cv,Buras:2021btx,Urquia-Calderon:2025wjx}.

The much lower sensitivity to exotic tau decays involving invisible particles, compared to fully visible final states, is generic and has to do with the fact that the production of taus in experiments always involves additional neutrinos (e.g., in $e^+e^-\to \tau^+ \tau^-$ the neutrino is from the decay of ``the other tau''). 
This has important consequences for searches. 
For instance, while, the experimental signature for $\tau\to \mu a$ decay  is a mono-energetic muon in the tau rest-frame, the tau rest-frame can be  reconstructed only approximately. 
The exotic two-body $\tau\to \mu a$ decay thus appears in the detector closer in shape to the irreducible SM background, the three body decay $\tau \to \mu \nu \bar \nu$, making the search more challenging. 

In this work we explore a much larger set of light new physics models that lead to exotic tau decays of the second type: the multi-lepton signatures. 
To some extent the new physics models that lead to multi-lepton signatures in tau decays are closely related to similar new physics models that result in rare multi-electron decays of muons, a number of which have already been discussed in the literature~\cite{Hostert:2023gpk,Heeck:2025jfs,Greljo:2025ljr}. 
However, often there are phenomenologically important differences between the two cases, which go beyond merely swapping taus with muons in the initial state. Most importantly, since tau is much heavier, the dark sector particles are kinematically allowed to decay to many different final states. For instance, a kinetically mixed dark photon with mass well above the muon threshold  predominantly decays to hadrons, while decays to muons and electrons are subleading. As we will see on concrete examples, it may thus be worthwhile to perform searches with simultaneous bump hunt searches in several  related final states.  

The paper is organized as follows. In \cref{sec:multilep_models} we first discuss general features of exotic multi-lepton decays, working within a simplified model,  and then introduce five benchmark models in \cref{sec:benchmark:models}. Current constrains are collected in \cref{sec:constraints}, while a brief discussion of experimental prospects can be found in \cref{sec:exp:prospects}, followed by conclusions in \cref{sec:conclusions}.

\section{Multi-lepton signatures} 
\label{sec:multilep_models}

\begin{figure}[t!]
    \centering\includegraphics[width=0.39\textwidth]{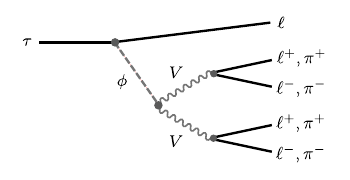}\\
    \includegraphics[width=0.36\textwidth]{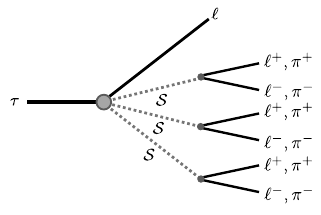}
    \caption{Dark cascaded decays of the type $\tau \to \lepton (2A) (2B)$ for $A,B \in \{\lepton,h\}$ (top) and $\tau \to \lepton (2A) (2B) (2C)$ for $A,B,C \in \{\lepton,h\}$ (bottom). 
    }\label{fig:tau_to_5_7}
\end{figure}

A multi-lepton signature can naturally arise in $\tau$ decays via decay chains in the dark sector that terminate with a decay back to SM leptons. 
We refer to these as dark cascades.
A token example is a two body $\tau\to \ell \phi$ decay, where $\phi$ is a light scalar or pseudoscalar that
decays to a pair of light dark vectors, $\phi\to VV$, followed by a subsequent decay of $V$
to SM lepton pairs, $V\to \ell^+\ell^-$.\footnote{We use $\phi$ for generic scalars or pseudo-scalars and $V$ for generic vector particles. 
When talking about specific model realizations, we distinguish dark photons from other realizations of vector particles by using $A'$ as opposed to $Z'$.
} The end result is a $\tau \to 5\ell$ decay, see \cref{fig:tau_to_5_7} (top). 

The form of various couplings, and the possible decay channels at each step, depend on the details of the new physics model, for which we give several examples below. 
However, the generic low energy phenomenology is already captured quite well by a simple
effective interaction Lagrangian of the form
\be\label{eq:generalL}
\mathcal{L}_{\text{eff}} \supset g_V V_\mu J^\mu + \frac{\mu_V}{2}\, \phi\, V_\mu V^\mu + (c_{ij}\, \phi \,\bar{\ell}_{L,i} \ell_{R,j} + \text{h.c.}),
\ee 
where we assumed that at tree-level $\phi$ interacts exclusively with SM leptons and the dark vector.
Depending on the UV completion, $g_V$ and $\mu_V$ may be correlated, though in the effective description in \cref{eq:generalL} they are treated as independent.
The SM charged lepton current is given by,
\begin{equation}\label{eq:V_current}
J^\mu = \bar{\psi}_L \gamma^\mu Q_L \psi_L + \bar{\psi}_R \gamma^\mu Q_R \psi_R,
\end{equation}
where $Q_{L,R}$ are $3\times 3$ hermitian matrices in flavor space with elements $[Q_{L/R}]_{ij} \equiv q_{ij}^{L/R}$, and $\psi_{L/R}=(e_{L/R},\mu_{L/R},\tau_{L/R})$ a vector of SM charged leptons. 
In general, $Q_{L,R}$ have both diagonal and off-diagonal entries. The off-diagonal entries $q^{L/R}_{i3}$ induce $\tau\to \ell_i V$ decays, $i=1,2$, with partial decay widths given by
\begin{equation}\label{eq:partialwidthtautoVmu}
\begin{split}
\Gamma(\tau \to \ell_i V) &= \frac{g_V^2 m_\tau}{32\pi}\lambda^{1/2}(1,r_{\lepton_i}^2, r_V^2)
\biggr[\left(|q_{i3}^L|^2+|q_{i3}^R|^2\right)
\\
&\times  \Big(1+r_{\ell_i}^2-2r_V^2 + (1-r_{\ell_i}^2)^2/{r_V^2}\Big)
\\
&-12\Re ({q^L_{i3}}^* q^R_{i3}) r_{\ell_i} \biggr],
\end{split}
\end{equation}
where $r_{\ell_i}=m_{\ell_i}/m_\tau$, $r_V=m_V/m_\tau$, and $\lambda(a,b,c) = a^2+b^2+c^2 - 2ab -2ac - 2bc$ is the K\"all\'en function.

Similarly, the Yukawa-like interaction, $c_{ij}$ in \cref{eq:generalL}, has in general both diagonal and off-diagonal entries. Note that in the UV complete theories, above electroweak symmetry breaking scale, these interactions are due to higher dimension operators, unless $\phi$ is part of an electroweak doublet. For a scalar $\phi$ with mass $m_\phi < m_\tau - m_{\ell_i}$, 
the $c_{3i,i3}$ couplings induce $\tau \to \ell_i \phi$ decays. 
The corresponding partial decay widths are
\begin{equation}
\label{eq:partialwidthtueatophimu}
\begin{aligned}
&\Gamma(\tau \to \ell_i \phi) = \frac{m_\tau}{32\pi}\lambda^{1/2}(1,r_{\lepton_i}^2,r_\phi^2)
\\
&\times
\left[(\vert c_{i3}\vert^2 + \vert c_{3i}\vert^2)(1+r_{\lepton_i}^2 - r_\phi^2) + 4\Re(c_{i3}c_{3i}) r_{\lepton_i}\right],
\end{aligned}
\end{equation}
where 
$r_\phi=m_\phi/m_\tau$. 

The experimental sensitivity for  $\mathcal{B}(\tau \to \ell_i\phi)$ depends on how $\phi$ decays.  More often than not, the dominant decay of $\phi$ is to a pair of dark vectors, with partial width given by
\be\label{eq:phitoAA}
\Gamma(\phi \to VV) = \frac{m_\phi^3  \mu_V^2}{128\pi m_V^4}(1-4r_{V/\phi}^2 + 12r_{V/\phi}^4)\sqrt{1-4r_{V/\phi}^2}~,
\ee
where $r_{V/\phi}=m_V/m_\phi$.
The $\phi$ may also decay to a pair of SM fermions.
For off-diagonal couplings the partial width is given by
\begin{equation}\label{eq:phitof1f2}
\begin{aligned}
    &\Gamma(\phi \to f_i f_j) 
= \frac{m_\phi}{8\pi}\lambda^{1/2}(1,r_i^2,r_j^2) \\ 
&\times \left[(\vert c_{ij}\vert^2 + \vert c_{ji}\vert^2) (1-r_i^2 - r_j^2) - 4 \mathrm{Re}(c_{ij}c_{ji}) r_i r_j\right],
\end{aligned}
\end{equation}
where $r_i=m_i/m_\phi$ and we have summed over the two possible charge assignments $\Gamma(\phi \to f_i f_j) = \Gamma(\phi \to f_i^- f_j^+) + \Gamma(\phi \to f_i^+ f_j^-)$. 
For flavor diagonal decays, assuming real $c_{ii}$, we have
\begin{equation}
    \Gamma(\phi \to f^+_i f^-_i) = \frac{c_{ii}^2 m_\phi}{8\pi} \left( 1-4r_i^2 \right)^{3/2}.
\end{equation}

The final step in the cascade decay is the decay of the dark vector to SM states. 
We will mainly be interested in decays to charged fermions, $f$, with charge $q$, for which the partial width is
\begin{align}\label{eq:partialwidthZprimetoff}
    \Gamma_{V \to f_i \bar{f}_j} &=  \frac{g_V^2 m_{V} }{48\pi}  \lambda^{1/2}(1,z_i^2,z_j^2)
    \\\nonumber
    &\times \Big[ \left(|q^L_{ij}|^2 + |q^R_{ij}|^2\right)\left(2 - z_i^2 - z_j^2 - (z_i^2 - z_j^2)^2\right) \\\nonumber
    &\qquad + 12 z_i z_j \Re({q^L_{ij}}^* q^R_{ij})\Big]~,
\end{align}
with $z_i = m_i/m_V$.  In some models there will be additional decay modes to neutrinos, $V\to \nu \bar \nu$. 
Additionally, in certain models, $V$ can also decay to light mesons. We give the corresponding decay widths below, see \cref{eq:Atopipi}.

In the numerical examples, we will use as a benchmark mass spectrum
\begin{equation}
\label{eq:benchmark_onshell}
    m_\phi = 1 \,\mathrm{GeV}, \quad m_{V} = 300 \,\text{MeV}.
\end{equation}
The masses of $\phi$ and $V$ were chosen such that in tau decays both $\phi$ and $V$ can be produced on-shell, and then decay promptly for the coupling ranges we are interested in.
In the prompt, on-shell regime, we therefore have 
\begin{equation}\begin{aligned}
    &\Gamma(\tau \to \ell_i \ell^+_j\ell^-_j\ell^+_k \ell^-_k) \simeq \\
    &\quad (2-\delta_{jk}) \Gamma(\tau \to \ell_i \phi) \times \mathcal{B}(V \to 2\ell_j) \times \mathcal{B}(V \to 2\ell_k),
\end{aligned}\end{equation}
where we assumed $\mathcal{B}(\phi \to 2V) \approx 1$.
In this regime, a measurement of the partial decay width $\Gamma(\tau \to \ell_i \ell^+_j\ell^-_j\ell^+_k \ell^-_k)$ gives a direct determination of $c_{3i}^2$, see \cref{eq:partialwidthtueatophimu}.

Above electroweak symmetry breaking scale, the effective coupling of $\phi$ to SM charged leptons, \cref{eq:generalL}, must arise from a higher dimension operator.  For the models we consider below the coupling can arise at dimension $5$ or $6$.  The translation of bounds on $c_{ij}$ to bounds on the scale of this higher dimension operator differs between the two cases. 

If the coupling of $\phi$ to SM charged leptons in \cref{eq:generalL} is due to a dimension-5 effective operator, the flavor violating interaction takes the form
\be\label{eq:generalL:dim5}
\mathcal{L}^{(5)} \supset  \frac{c^{(5)}_{ij}}{\Lambda_5} \Phi  (\bar{L}_{L,i} \ell_{R,j}) H + \text{h.c.},
\ee
where $\Phi$ denotes the full scalar field prior to symmetry breaking.
After electroweak and $U(1)'$ symmetry breaking, writing $\Phi = (v' + \phi)/\sqrt{2}$ and $H^0 = (v + h)/\sqrt{2}$, the effective coupling is parametrically given by
\beq
c\sim c^{(5)} \frac{v}{2 \Lambda_5} ~,
\eeq
where $v$ is the vacuum expectation value (VEV) of the SM Higgs and we have suppressed the flavor indices which will depend upon the rotation matrices necessary to diagonalize the charged lepton masses.
On the other hand, if the effective coupling is due to a dimension $6$ effective operator of the form 
\be\label{eq:generalL:dim6}
\mathcal{L}^{(6)} \supset  \frac{c^{(6)}_{ij}}{\Lambda_6^2} |\Phi|^2  (\bar{L}_{L,i} \ell_{R,j}) H + \text{h.c.},
\ee 
then the effective coupling is 
\beq
c \sim c^{(6)} \frac{v' v}{\sqrt{2}\Lambda_6^2},
\eeq
where $v'$ is the VEV of a light scalar field. 

These scalings mean that a bound on the tau branching fraction scales as 
\beq
\begin{split}
\label{eq:Br:tau:ell:phi:example}
\mathcal{B}(\tau \to \ell_i\phi)&\simeq 10^{-8} \cdot \biggr(\frac{ c_{i3}}{10^{-9}}\biggr)^2
\\
&\simeq 10^{-8} \cdot \left(c_{i3}^{(5)}\right)^2 \cdot \biggr(\frac{10^{11}\,\text{GeV}}{\Lambda_{5}}\biggr)^2
\\
&\simeq 10^{-8} \cdot  \left(c_{i3}^{(6)}\right)^2 \cdot \biggr(\frac{10^{6}\,\text{GeV}}{\Lambda_{6}}\biggr)^4 \cdot \biggr(\frac{v'}{10\,\text{GeV}}\biggr)^2~.
\end{split}
\eeq
For simplicity we have considered the situation with only one non-zero $c_{ij}$.  
From \cref{eq:Br:tau:ell:phi:example} we see that whether the interaction of $\phi$ to the SM fields is due to a dimension 5 operator or due a dimension 6 one, the new physics scale that can in principle be probed in rare tau decays lies well above the electroweak symmetry breaking scale and also well above the direct reach in the colliders.

One important distinction between dimension-5 and dimension-6 operators is the correlation between $\tau \to \ell_i V$ and $\tau \to \ell_i \phi$.
If the mass of $V$ comes from the expectation value of the complex scalar that contains $\phi$, the dimension-5 operators will typically predict 
\begin{equation}\label{eq:ratio_V_over_phi_taudecay}
    \frac{\mathcal{B}(\tau \to \ell_i V)}{\mathcal{B}(\tau \to \ell_i \phi) } \simeq 1
\end{equation}
when both $\phi$ and $V$ are light with respect to $m_\tau$ as a direct consequence of the Goldstone boson equivalence theorem.
This can be understood as the fact that the phase of the complex scalar is allowed to couple to the flavor-violating current.
For dimension-6 operators, this is not the case and we typically expect the ratio in \cref{eq:ratio_V_over_phi_taudecay} to be small or to vanish altogether.
In practice, when $V\to \ell^+ \ell^-$ and $\phi \to VV$ dominate, the ratio in \cref{eq:ratio_V_over_phi_taudecay} controls the hierarchy between three-lepton and five-lepton tau decays.


\section{Benchmark models}
\label{sec:benchmark:models}

Next, let us turn to specific realizations of this general dark cascade phenomenology. 
We will discuss five examples: a kinetically mixed photon (\cref{sec:Model:I:kin:mix}); a gauged $U(1)$ corresponding to a difference of lepton numbers between two families, e.g.,~$L_\mu-L_\tau$ (\cref{sec:Model:II:gauged:LX-LY}); a gauged $U(1)$ that is a linear combination of all three lepton family numbers (\cref{sec:model:III:gauge:all}); a chiral model (\cref{sec:Model:IV:chiral:model}); and a flavor protected scalar (\cref{sec:model:V:flavor:protected:scalar}). 

The most important decay modes for each of the representative models are summarized in \cref{tab:allmodes}. 
From this table it is clear that if a signal in any mode is found, searching in additional leptonic or hadronic modes will allow for good model discrimination. 

\begin{table*}[ht]
\centering
\renewcommand{\arraystretch}{1.3}
\begin{tabular}{c l c c c c c c c c}
Model & & &  \hyperref[sec:Model:I:kin:mix]{I} &   \hyperref[sec:Model:II:gauged:LX-LY]{IIa} &   \hyperref[sec:Model:II:gauged:LX-LY]{IIb} &   \hyperref[sec:model:III:gauge:all]{III} &   \hyperref[sec:Model:IV:chiral:model]{IVa} &   \hyperref[sec:Model:IV:chiral:model]{IVb} &   \hyperref[sec:model:V:flavor:protected:scalar]{V}\\
 \hline
\hline
 &  &  & ~Seclud. $U(1)$~ & $U(1)_{L_\mu - L_\tau}$ & $U(1)_{L_\mu - L_\tau}$ & $U(1)_{L_\mu + L_e - 2L_\tau}$ & ~Chiral~  & ~Chiral~  & Flav. prot. \\
 $\slashed{E}$ & $\tau$ decay  & ~Search~ & Kinetic mix. & $Q_\Phi = +1$ & $Q_\Phi = +2$ & $Q_\Phi = -3$ & $U(1)_\mu$ & $U(1)_e$ & scalar\\
\hline
& $\tau \to \muon{3\mu}$ & \cellcolor{green!25}\cite{Belle-II:2024sce} & \color{gray}{(\ding{51})} & & \ding{51} & \ding{51} & \ding{51} & & \color{gray}{(\ding{51})} \\
$\slashed{E} = 0$ & $\tau \to \muon{2\mu} \, \el{e}$ & \cellcolor{green!25}\cite{Hayasaka:2010np} & \color{gray}{(\ding{51})} & \ding{51} & &  \ding{51} & & & \\
3 leptons & $\tau \to \muon{\mu} \, \el{2e}$ & \cellcolor{green!25}\cite{Hayasaka:2010np} & \color{gray}{(\ding{51})} & & &  \ding{51} & & & \color{gray}{(\ding{51})} \\
& $\tau \to \el{3e}$ & \cellcolor{green!25}\cite{Hayasaka:2010np} & \color{gray}{(\ding{51})} & & &  \ding{51} & & \ding{51} & \\
\hline
$\slashed{E} > 0$ & $\tau \to \muon{\mu} \, 2\nu$ &\cellcolor{green!25}\cite{Belle-II:2022heu,Belle:2025bpu} & & & \ding{51} &  \ding{51} & \ding{51} & & \\
3 leptons & $\tau \to \el{e} \, 2\nu$ & \cellcolor{green!25}\cite{Belle-II:2022heu,Belle:2025bpu} & & \ding{51} & &  \ding{51}& & \ding{51} & \\
\hline
& $\tau \to \muon{5\mu}$ &  \cellcolor{red!25}\cite{Overhoff:2020}$^\mathbf{\ast}$ & \ding{51} & & \ding{51} & \ding{51} & \ding{51} & & \color{gray}{(\ding{51})} \\
& $\tau \to \muon{4\mu}\, \el{e}$ & \cellcolor{red!25}\ding{55} & \ding{51} & \ding{51} &  & \ding{51} & & &  \color{gray}{(\ding{51})} \\
$\slashed{E} = 0$ & $\tau \to \muon{3\mu} \,  \el{2e}$ & \cellcolor{red!25}\ding{55} & \ding{51} &  &  & \ding{51} & & &  \color{gray}{(\ding{51})} \\
5 leptons & $\tau \to \muon{2\mu} \, \el{3e}$ & \cellcolor{red!25}\ding{55} & \ding{51} &  &  & \ding{51} & & &  \color{gray}{(\ding{51})}\\
& $\tau \to \muon{\mu} \, \el{4e}$ & \cellcolor{red!25}\ding{55} & \ding{51} &  &  & \ding{51} & & &  \color{gray}{(\ding{51})}\\
& $\tau \to \el{5e}$ & \cellcolor{red!25}\ding{55} & \ding{51} &  &  & \ding{51} & & \ding{51} &  \color{gray}{(\ding{51})} \\
\hline
& $\tau \to \muon{3\mu} \, \had{2\pi}$ & \cellcolor{red!25}\ding{55} & \ding{51} &  &  & & & & \\
& $\tau \to \muon{2\mu}\, \el{e}\, \had{2\pi}$ & \cellcolor{red!25}\ding{55} & \ding{51} &  &  & & & & \\
$\slashed{E} = 0$ & $\tau \to \muon{\mu} \, \el{2e}\, \had{2\pi}$ & \cellcolor{red!25}\ding{55} & \ding{51} &  & & & & & \\
hadrons & $\tau \to \muon{\mu} \, \had{4\pi}$ & \cellcolor{red!25}\ding{55} & \ding{51} &  &  & & & & \\
& $\tau \to \el{3e} \,  \had{2\pi}$ & \cellcolor{red!25}\ding{55} & \ding{51} &  &  & & & & \\
& $\tau \to \el{e} \, \had{4\pi}$ & \cellcolor{red!25}\ding{55} & \ding{51} &  &  & & & & \\
\hline
& $\tau \to \muon{3\mu} \, 2\nu$ & \cellcolor{red!25}\ding{55} &  & & \ding{51} & \ding{51} & \color{gray}{(\ding{51})} & & \\
& $\tau \to \muon{2\mu} \, \el{e}\, 2\nu$ & \cellcolor{green!25}\cite{CLEO:1995azm} & & \ding{51} &  & \ding{51} & & & \\
$\slashed{E} > 0$ & $\tau \to \muon{\mu} \, \el{2e}\,2\nu$ & \cellcolor{red!25}\ding{55} &  &  &  & \ding{51} & &  & \\
5 leptons& $\tau \to \muon{\mu} \, 4\nu$ & \cellcolor{red!25}\ding{55} &  & & \ding{51} & \ding{51} & \color{gray}{(\ding{51})} & & \\
& $\tau \to \el{3e} \, 2\nu$ & \cellcolor{green!25}\cite{CLEO:1995azm} &  &  & & \ding{51} & & \color{gray}{(\ding{51})} & \\
& $\tau \to \el{e} \, 4\nu$ & \cellcolor{red!25}\ding{55} & & \ding{51} &  & \ding{51} &  & \color{gray}{(\ding{51})} & \\
\hline
& $\tau \to \muon{7\mu}$ & \cellcolor{red!25}\ding{55}
&  &  &  &  &  &  & \ding{51} \\
& $\tau \to \muon{6\mu\,}\, \el{e}$ & \cellcolor{red!25}\ding{55}
&  &  &  &  &  &  & \ding{51} \\
$\slashed{E} = 0$ & $\tau \to \muon{5\mu} \,  \el{2e}$ & \cellcolor{red!25}\ding{55}
&  &  &  &  &  &  & \ding{51} \\
7 leptons & $\tau \to \muon{4\mu} \, \el{3e}$ & \cellcolor{red!25}\ding{55}
&  &  &  &  &  &  & \ding{51} \\
& $\tau \to \muon{3\mu} \, \el{4e}$ & \cellcolor{red!25}\ding{55}
&  &  &  &  &  &  & \ding{51} \\
& $\tau \to \muon{2\mu} \el{5e}$ & \cellcolor{red!25}\ding{55}
&  &  &  &  &  &  & \ding{51} \\
& $\tau \to \muon{\mu} \el{6e}$ & \cellcolor{red!25}\ding{55}
&  &  &  &  &  &  & \ding{51} \\
& $\tau \to \el{7e}$ & \cellcolor{red!25}\ding{55}
&  &  &  &  &  &  & \ding{51} \\
\hline
\hline
\end{tabular}
\caption{Master atlas for three-, five-, and seven-body $\tau$ decays into leptons and hadrons ($\nu$ denotes both SM and sterile neutrinos). A checkmark indicates which model contributes at tree level, while greyed out checkmarks {\color{gray}{(\ding{51})}} denote that the signature is possible but is suppressed by higher powers of the new physics scale $\Lambda$ in benchmark models. The third column indicates whether there has been an experimental search for the given decay channel: \colorbox{green!25}{green} indicates an existing search with the corresponding reference, while \colorbox{red!25}{red} indicates no dedicated search has been performed. The $\ast$ denotes a preliminary LHCb sensitivity study. 
Near the $\omega$ resonance ($m_V \approx 782\,$MeV), $V \to \pi^+\pi^-\pi^0$ can also contribute in Model I, giving additional hadronic channels, i.e., replacing $2\pi \to 3\pi$ and $4\pi \to 5\pi+6\pi$ in the above list of final states, see the inset in \cref{fig:model1_BRs}.
}
\label{tab:allmodes}
\end{table*}

\begin{figure}[t!]
    \centering
    \includegraphics[width=0.49\textwidth]{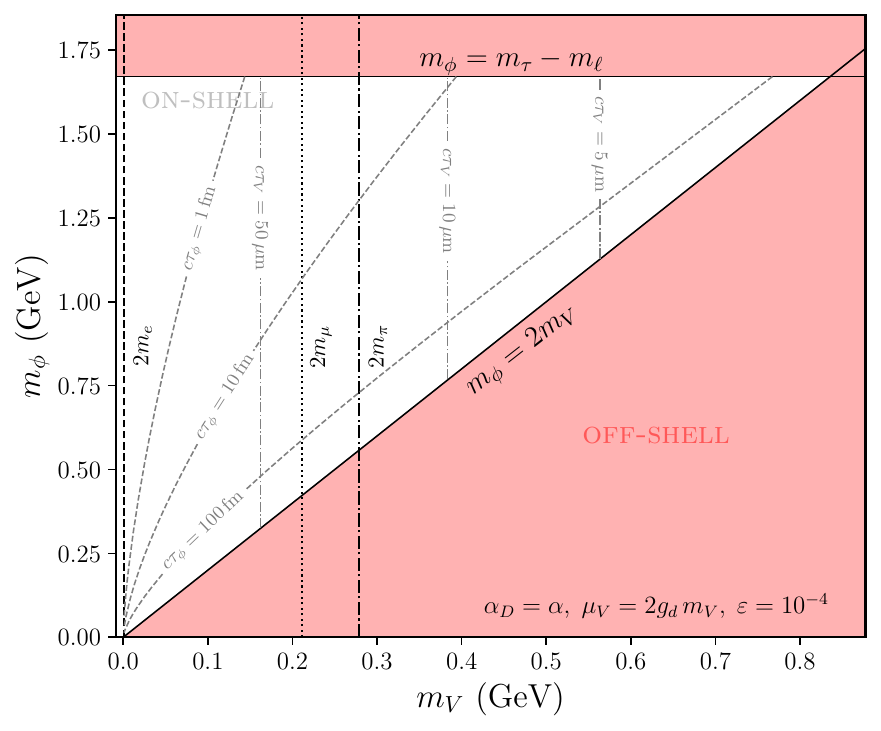}
    \caption{Viable parameter space for prompt and on-shell light new physics sourcing $\tau \to$ multi-lepton cascade decay signatures assuming $\varepsilon = 10^{-4}, \alpha_d = \alpha$.
    }\label{fig:mZp_mphi}
\end{figure}

\subsection{Model I: kinetic mixing}
\label{sec:Model:I:kin:mix}

We start with a particularly minimal extension of the SM: the dark photon~\cite{Holdom:1985ag,Caputo:2021eaa}, now denoted by $A'$ (that is, $V = A'$ in what follows).
We assume the mass of $A'$ is generated by a light dark Higgs $\Phi$.
It is this scalar particle that, through non-renormalizable couplings, can initiate LFV decays of SM leptons.
The radial mode of the dark Higgs, $\phi$, initiates the chain of decays via $\tau \to \ell_i (\phi \to A' A')$ in much the same way that it did in the $\mu\to e (\phi \to A'A')$ decay studied in Ref.~\cite{Hostert:2023gpk}. 
Because $A'$ only couples to the SM via kinetic mixing, the decay chain can only terminate on multi-lepton and multi-pion final states, without any missing energy.
Examples of such dark cascade signatures include
$\tau \to 5\mu$, $\tau \to 2\pi 3\mu$, $\tau \to 4\pi \mu$, $\tau \to 2e 3\mu$ (and equivalent decays with $\mu\leftrightarrow e$), see \cref{tab:allmodes}.

We now turn to a quantitative formulation of the above discussion.
The $U(1)'$ part of the Lagrangian is, for energies well below the electroweak scale, given by 
\begin{equation}\label{eq:phi_lag}
    \mathcal{L}_{U(1)'} = - \frac{1}{4}F_{\mu\nu}^{\prime 2} - \frac{\varepsilon}{2} F'_{\mu\nu} \hat{F}^{\mu\nu} + |D_\mu \Phi|^2 - V(\Phi),
\end{equation}
where $D_\mu \equiv \partial_\mu - i g_d {A}'_\mu$, $\varepsilon$ is the kinetic mixing parameter. The dark scalar potential is\footnote{Above the electroweak scale, the potential in general also includes a quartic coupling to the SM Higgs boson, $V(\Phi)\supset \lambda_{\Phi H} |\Phi|^2 |H|^2$. Below electroweak symmetry breaking this contributes to $\Phi$ mass term.}
\begin{equation}\label{eq:scalar_potential}
    V(\Phi) = -\mu^2 |\Phi|^2 + \lambda |\Phi|^4,
\end{equation}
The $U(1)'$ is spontaneously  broken once the field $\Phi$ obtains a nonzero VEV,
$\langle \Phi \rangle \equiv v'/\sqrt2 \ll \langle H \rangle \equiv v/\sqrt2$. The kinetic mixing term in Eq.~\eqref{eq:phi_lag} can be removed via the field redefinition $\hat{A}_\mu \to A_\mu - \varepsilon A_\mu'$, giving canonically normalized $A, A'$, and inducing millicharged couplings of SM leptons and hadrons to $A'$. The mass of $A'$ is proportional to $U(1)'$ symmetry breaking, and is given by $m_{A'} = g_d v'$
assuming that $\Phi$ has the $U(1)'$ charge $+1$.

The LFV is induced in this model via non-renormalizable interactions between $\Phi$ and SM fields. 
Generically, the first $SU(2)_L$-invariant LFV operators appear at dimension-5, though these would require some of the SM fields to be charged under $U(1)'$, which we assume not to be the case for the dark photon.
Therefore, the first $U(1)'$-invariant LFV operators appear at dimension-6,
\begin{equation}\label{eq:LFV_lag_I}
    -\mathcal{L}_{\text{LFV},{\rm I}} = \frac{\kappa_{ij} }{\Lambda^2} |\Phi|^2 \bar{L}_{i} H  \ell_{Rj} + \text{h.c.},
\end{equation}
where $L = (L_1, L_2, L_3)^T$ and $\ell_R = (e_{R}, \mu_{R},\tau_R)^T$ are vectors in flavor space with the left and right-handed lepton doublets, respectively, where $L_1=(\nu_e, e_L)$, $L_2=(\nu_\mu, \mu_L)$, $L_3=(\nu_\tau, \tau_L)$, and $\kappa_{ij}$ is a matrix of Wilson coefficients in flavor space.

After both $\Phi$ and $H$ obtain VEVs, the leptonic mass matrix receives two contributions: one from the dimension-6 operator in \cref{eq:LFV_lag_I} and another from the SM Yukawa couplings,
\begin{equation}
\label{eq:SM:Yukawa}
    \mathcal{L}_{\rm SM} \supset -y_{ij}\bar{L}_i H \ell_{Rj} + {\rm h.c.},
\end{equation}
where $y_{ij}$ is the Yukawa matrix in flavor space.
The charged lepton masses are thus given by,
\begin{equation}
\label{eq:Model:I:mass:matrix}
    \text{diag}(m_e, m_\mu, m_\tau) = U_L^\dagger\left(y + \kappa\frac{v'^2 }{2\Lambda^2}\right)U_R \frac{v}{\sqrt{2}},
\end{equation}
where $U_{L,R}$ are the unitary matrices that diagonalize the mass matrix.
After transforming to the mass basis, the low energy coefficients in \cref{eq:generalL} are given by  
\begin{equation}
    c_{ij}= -(U_L^\dagger \kappa \, U_R)_{ij}\frac{v' v}{\sqrt{2}\Lambda^2}, \,\,\mu_V = 2 g_d m_{A'}, \text{ and } g_V = e \varepsilon.
\end{equation}
Note that because the dark photon inherits the same flavor-conserving structure as the SM photon, $Q_L = Q_R = \mathbf{1}$, all interactions between the dark photon and the SM leptons are flavor conserving.

\begin{figure*}[t]
    \centering
    \includegraphics[width=0.85\textwidth]{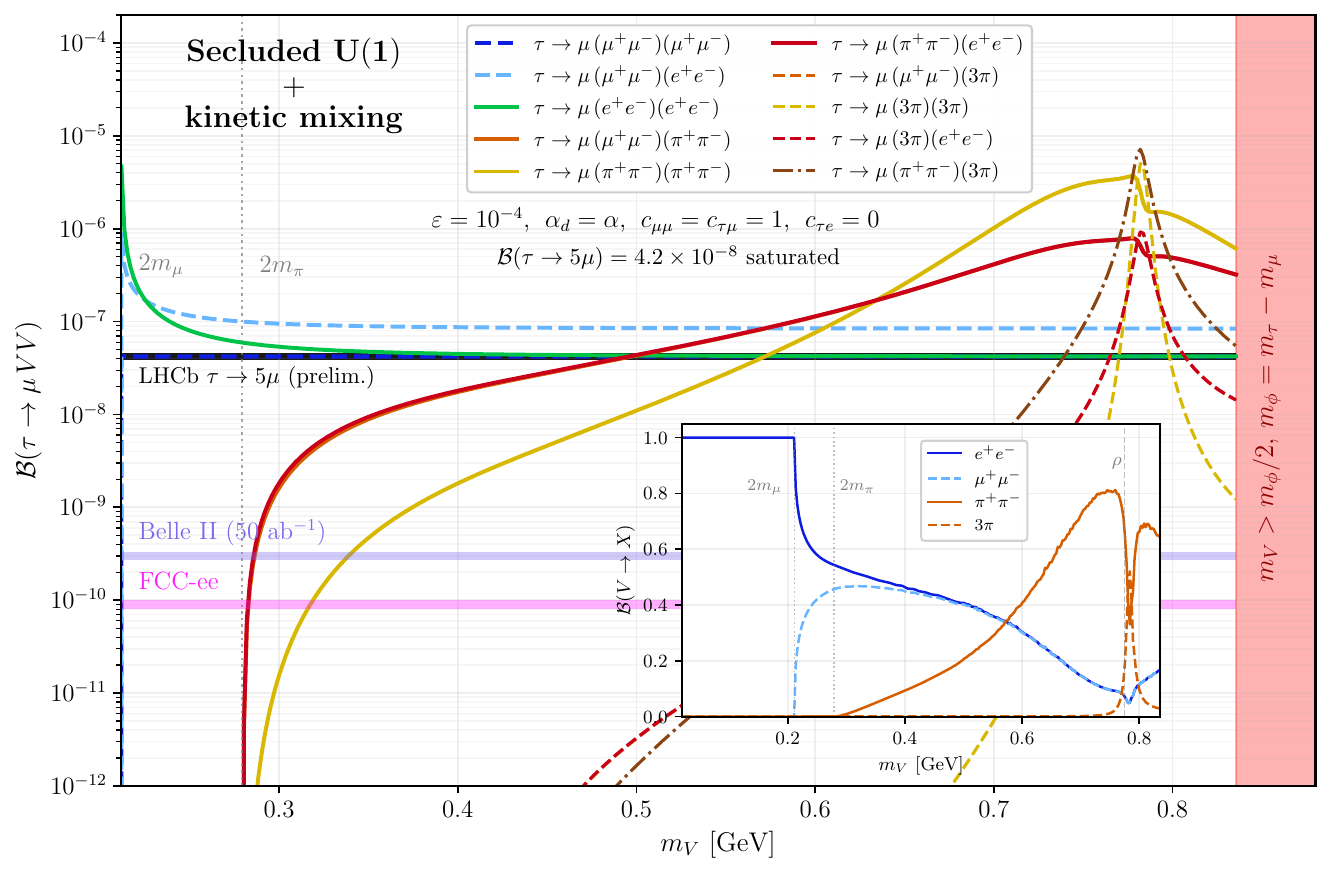}
    \caption{Predicted branching ratios for $\tau \to \mu\phi \to \mu VV$ cascade decays in Model~I (kinetic mixing) as a function of the dark photon mass $m_V$. 
    The scalar mass is set to its maximal value, $m_\phi = m_\tau - m_\mu \approx 1.67$~GeV, for which both $\phi$ and $V$ decay promptly and on-shell for benchmark couplings $\varepsilon = 10^{-4}$ and $\alpha_D = \alpha$. 
    For simplicity, we assume only $\kappa_{\tau \mu}$ is non-zero, the resulting Wilson coefficient $c_{\tau\mu}$ is then determined at each $m_V$ by saturating the preliminary LHCb $\tau \to 5\mu$ sensitivity of $4.2 \times 10^{-8}$ (black dashed horizontal line),
    pinning the $\tau \to 5\mu$ channel to this value across all masses.
    Consequently, any channel with predicted branching ratio above this line would be observable with comparable experimental sensitivity. 
    The solid blue curve shows $\tau \to 5\mu$, while dashed and dot-dashed curves show mixed leptonic ($3\mu 2e$, $\mu 4e$) states. Hadronic channels are separated into $2\pi$ ($\rho^0$) and $3\pi$ ($\omega/\phi$) contributions using the VMD decomposition of~\cite{Ilten:2018crw}; solid lines show the $\pi^+\pi^-$ channels ($3\mu\,2\pi$, $\mu\,4\pi$, $\mu\,2\pi\,2e$) and dashed lines the corresponding $3\pi$ channels. 
    The red shaded region at $m_V > m_\phi/2$ indicates where $\phi \to VV$ becomes kinematically forbidden. 
    Horizontal lines show the projected Belle~II reach (purple band, $\sim 3 \times 10^{-10}$) and FCC-ee sensitivity (magenta band, $\sim 9\times 10^{-11}$), obtained by rescaling current upper limits by the expected increase in tau statistics (see \cref{sec:exp:prospects}). 
    \textit{Inset:} Dark photon branching ratios $\mathcal{B}(V \to X)$ for $X = e^+e^-$, $\mu^+\mu^-$, $\pi^+\pi^-$ and $\pi^+\pi^-\pi^0$,
    computed using the experimentally measured $R$-ratio. Near the $\rho$ resonance ($m_V \approx 775$~MeV), hadronic channels dominate, causing $\tau \to \mu(n\pi)$ final states to dominate over the purely leptonic channels by an order of magnitude.
    }
    \label{fig:model1_BRs}
\end{figure*}

For $2 m_e < m_{A'} < 2 m_{\pi}$, dark photons will decay exclusively through the dilepton channel, with partial width given by \cref{eq:partialwidthZprimetoff}.
For $m_{A'} > 2m_{\pi}$, $A'$ will also decay to hadrons with partial width
\begin{equation}\label{eq:Atopipi}
    \Gamma_{A' \to {\rm hadrons}} = \Gamma_{A' \to \mu\mu} \times R(s = m_{A'}^2),
\end{equation}
where $R \equiv \sigma_{e^+e^- \to \text{hadrons}} / \sigma_{e^+e^- \to \mu^+ \mu^-}$.
Note that the only hadronic  decay channels of $A'$ that are kinematically allowed are into pions. Here, the $A'\to \pi^+\pi^-$ is the dominant one, except when $A'$ mass is almost degenerate with $\omega$, in which case $A'\to \pi^+\pi^- \pi^0$ is the most important channel 
(in most of the discussion of this model we neglect this signature for brevity). The $A'\to 4\pi$ channel  is always subdominant.  
To separate these contributions in \cref{fig:model1_BRs}, we decompose the total $R$-ratio into exclusive hadronic channels using the vector meson dominance (VMD) form factors of~\cite{Ilten:2018crw}. 
The branching ratios for the main decay channels of $A'$, as a function of $m_{A'}$, are shown in (the inset of) \cref{fig:model1_BRs}.

For $m_\phi > 2 m_{A'}$, $\phi$ will decay predominantly to dark photons, $\phi \rightarrow 2 A'$, provided the dark couplings are sufficiently larger than the non-renormalizable LFV interactions involving first and second generation leptons, so as to dominate over $\phi \to \ell_i \ell_j$ decays. 
For the benchmark numerical values we use, 
\beq
\label{eq:benchmark:varepsilon}
\varepsilon = 10^{-4} \text{~and~}\alpha_d = \alpha,
\eeq
along with the usual benchmark mass spectrum, \cref{eq:benchmark_onshell},
\beq
\label{eq:benchmark}
m_{A'} = 300\text{\,MeV and~}m_\phi = 1\text{\,GeV}.
\eeq
For these benchmark values both $\phi$ and $A'$ decay promptly, well within the viable parameter space shown in \cref{fig:mZp_mphi}, 
\begin{align}
 c\tau_{A'} &\simeq 14 \,\mu\text{m} \left( \dfrac{10^{-4}}{\varepsilon} \right)^2 \left( \dfrac{300\,\text{MeV}}{m_{A'}} \right),
 \\
    c\tau_{\phi} &\simeq 33 \,\text{fm} \left( \frac{\alpha}{\alpha_d} \right) \left( \frac{1\,\text{GeV}}{m_\phi} \right)^3 \left( \frac{m_{A'}}{300 \text{\,MeV}} \right)^2.
\end{align}
The lifetime of $\phi$ is determined by its decay $\phi\to 2A'$ for $\vert c_{ij} \vert \lesssim 10^{-1}$, which is the situation we are interested in.  At the benchmark point the vector decays slightly more to electrons than muons and has a $\mathcal{O}(1\%)$ branching fraction to charged pions. 
The scalings in the above estimates ignore the masses of the final state particles in the phase space factors, see \cref{eq:phitoAA,eq:partialwidthZprimetoff}, for complete expressions. 

\begin{figure*}[t!]
    \centering
    \includegraphics[width=0.99\textwidth]{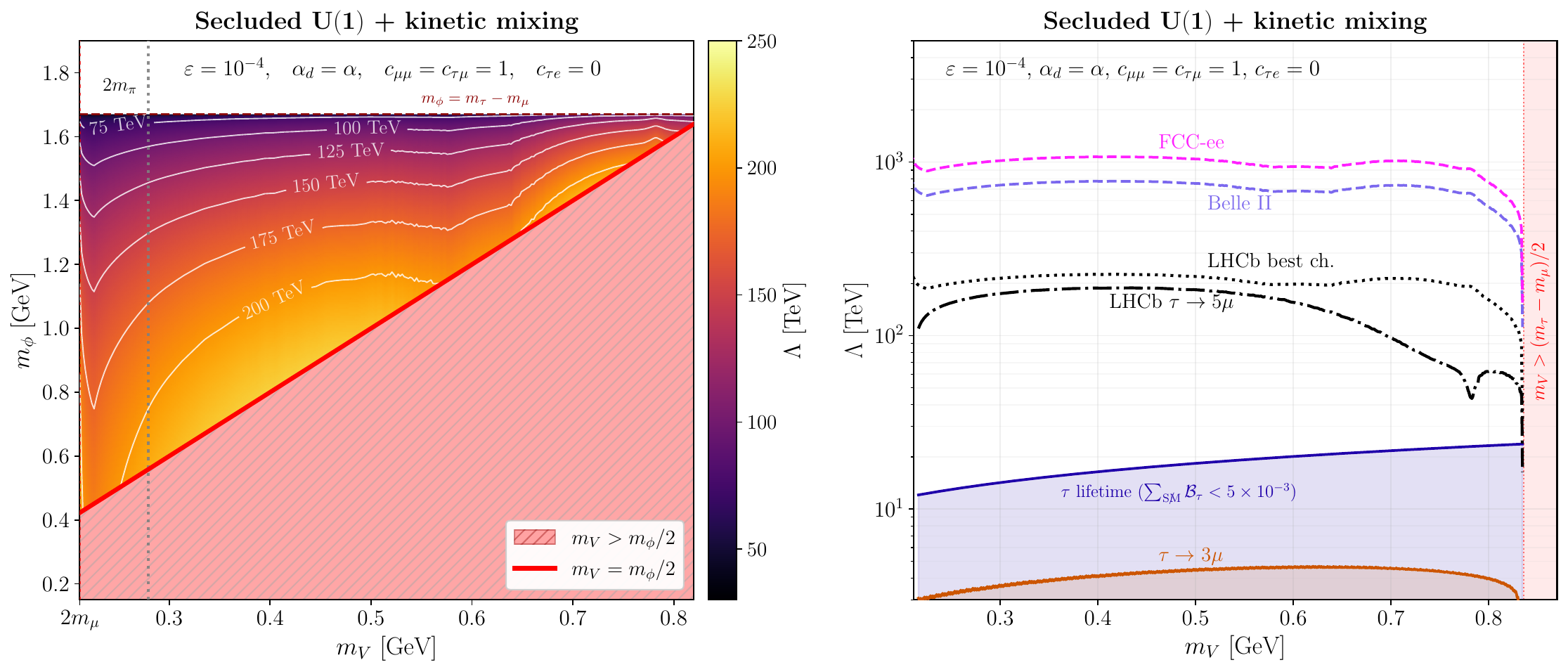}
    \caption{Sensitivity to the new physics scale $\Lambda$ in the secluded $U(1)'$ model with kinetic mixing (Model I). \textit{Left:} The NP scale $\Lambda$ reach of $\tau \to 5\mu$ corresponding to the LHCb preliminary sensitivity $\mathcal{B}(\tau\to 5\mu) < 4.2\times 10^{-8}$, shown in the $(m_V, m_\phi)$ plane. 
    The red shaded region is kinematically forbidden ($m_V > m_\phi/2$). \textit{Right:} Discovery reach as a function of $m_V$, maximized over $m_\phi$ for each $m_V$. The shaded regions show existing constraints from the $\tau$ lifetime (blue, $\sum \mathcal{B}_{\slashed{\rm SM}} < 5\times10^{-3}$, maximized over all kinematically allowed $m_\phi$) and from $\tau\to 3\mu$ (orange), the latter requiring a nonzero diagonal coupling $c_{\mu\mu} = c_{\tau\mu}$ (see text). All other curves are maximized over $m_\phi > 2m_V$, i.e., the on-shell cascade regime $\phi \to VV$.
    Solid and dashed curves show the $\Lambda$ reach of LHCb (preliminary, best channel and $5\mu$-only), 
    In both panels $\varepsilon = 10^{-4}$, $\alpha_d = \alpha$, $c_{\mu\mu} = c_{\tau\mu} = 1$, and $c_{\tau e} = 0$.
    }\label{fig:model1_Lambda}
\end{figure*}

As a benchmark we work directly with the low-energy couplings $c_{ij}$ in \cref{eq:generalL}, and take 
\beq
c_{\mu\mu} = c_{\mu\tau} =1,
\eeq
 with all other $c_{ij} = 0$. This is the minimal choice that induces both exotic five-body decays such as $\tau \to 5\mu$ (via the flavor-violating $c_{\mu\tau}$) and allows the $\tau \to 3\mu$ constraint to apply (which requires the flavor-diagonal $c_{\mu\mu}$). 
A more general flavor structure, with additional nonzero entries $c_{ij}$, would open further decay channels but does not change the qualitative conclusions regarding the discovery reach. 
The predicted branching ratios as a function of $m_V$ are shown in \cref{fig:model1_BRs}. For each value of $m_V$ the NP scale $\Lambda$ was set such that the $\tau \to 5\mu$ branching ratio equals the preliminary estimate for LHCb sensitivity, $\text{Br}(\tau \to 5\mu)=4.2 \times 10^{-8}$ \cite{Overhoff:2020}. Note that for larger $m_V$ it may be advantageous to search for mixed lepton-hadronic modes, such as $\tau \to 3\mu 2\pi$, $\tau \to \mu 4\pi$ or $\tau \to 3\mu 3\pi$ (if $V$ mass is close to $\omega$ meson), which are the dominant decay modes in that mass range.
The value of  $\Lambda$ that gives  $\text{Br}(\tau \to 5\mu)=4.2 \times 10^{-8}$, as a function of $m_\phi$ and $m_V$ is shown in the left panel in \cref{fig:model1_Lambda} (the shaded region denotes $m_V>m_\phi/2$, in which case the $\phi$ is off-shell, suppressing the $\tau\to 5\mu$ signal).

The right panel in \cref{fig:model1_Lambda} shows the projected sensitivity to $\Lambda$ at Belle II (with $50\,\text{ab}^{-1}$) and FCC-ee. These projections are obtained by rescaling the preliminary LHCb $\tau \to 5\mu$ sensitivity of $\mathcal{B} < 4.2 \times 10^{-8}$\cite{Overhoff:2020} by the expected increase in tau statistics at each facility, assuming comparable per-tau sensitivity (see \cref{sec:exp:prospects}). This gives projected sensitivities of $\mathcal{B} \sim 3 \times 10^{-10}$ for Belle~II and $\mathcal{B} \sim 8 \times 10^{-11}$ for FCC-ee, applied uniformly to all cascade final states. 
At each value of $m_V$, the reach corresponds to the ``best channel'', \textit{i.e.},~whichever cascade channel yields the strongest constraint. Combining several channels can lead to further improvements in the reach.

\subsection{Model II: gauged \texorpdfstring{$L_\alpha - L_\beta$}{L_X - L_Y}} 
\label{sec:Model:II:gauged:LX-LY}
Similar multilepton signatures arise in models with a local $U(1)'$ that gauges differences of family lepton numbers: $L_e - L_\mu, L_e - L_\tau$, or  $L_\mu - L_\tau$. The $U(1)'$ is spontaneously broken once the scalar $\Phi$, charged under $U(1)'$, obtains a VEV, which also gives mass to the $U(1)'$ gauge boson $V \equiv Z'$.
Each of the three $L_\alpha - L_\beta$ choices is anomaly-free, and can lead to similar signatures to those discussed  in \cref{sec:Model:I:kin:mix} for Model I, as we show below. 

For all three  $L_\alpha - L_\beta$ choices, the $U(1)'$ Lagrangian at low energy, below electroweak scale, is given by
\begin{equation}
\begin{split}
\label{eq:Lagr:Ui-j}
    \mathcal{L}_{U(1)_{\alpha-\beta}'} =& - \frac{1}{4}Z_{\mu\nu}^{\prime 2} - \frac{\varepsilon}{2} Z_{\mu\nu}' \hat{F}^{\mu\nu} + g' Z_\mu' J^\mu_{\alpha-\beta}
    \\
    & + |D_\mu \Phi|^2 - V(\Phi), 
\end{split}
\end{equation}
where $D_\mu \equiv \partial_\mu - i g' Q_\Phi Z'_\mu$, with $Q_\Phi$ the $L_\alpha-L_\beta$ charge of $\Phi$, $\varepsilon$ is the kinetic mixing parameter. The charged lepton current $J^\mu_{\alpha-\beta} $ is given by
\begin{equation}
\label{eq:Jmu:i-j}
\begin{split}
    J^\mu_{\alpha-\beta} &= \bar{L} \gamma^\mu Q_{\alpha-\beta} L + \bar{\ell}_R \gamma^\mu Q_{\alpha-\beta} \ell_R\\
    & = \bar{\ell}_\alpha \gamma^\mu \ell_\alpha - \bar{\ell}_\beta \gamma^\mu \ell_\beta + \bar{\nu}_\alpha \gamma^\mu P_L \nu_\alpha - \bar{\nu}_\beta \gamma^\mu P_L \nu_\beta,
\end{split}
\end{equation}
where $L = (L_1, L_2, L_3)^T, \ell_R = (e_{R}, \mu_{R}, \tau_R)^T$, and $Q_{\alpha-\beta}$ is the $U(1)'$ leptonic charge matrix. For example, in $L_\mu - L_\tau$ case, $Q_{\mu - \tau} = \text{diag}(0,+1, -1)$. The scalar potential $V(\Phi)$ in \cref{eq:Lagr:Ui-j} is taken to be the same as in Model I, see \cref{eq:scalar_potential}. 
The VEV $\langle \Phi \rangle \equiv v'/\sqrt2$ spontaneously breaks $U(1)'$, giving $Z'$ a mass $m_{Z'} = g'v' Q_\Phi$. 

The kinetic mixing term in Eq.~\eqref{eq:Lagr:Ui-j} can be removed via field redefinition $\hat{A}_\mu \to A_\mu - \varepsilon Z_\mu'$, giving canonically normalized $A$ and $Z'$. This also induces millicharged ${\mathcal O}(\varepsilon)$ couplings of SM quarks to $Z'$, and changes the couplings of  $Z'$ to the SM charged leptons by ${\mathcal O}(\varepsilon)$ away from the $J_{\alpha-\beta}^\mu$ current in \cref{eq:Jmu:i-j}. 
We will assume that $\varepsilon e\ll g'$ so that the $Z'$ couplings to leptons due to $L_\alpha-L_\beta$ current in \cref{eq:Jmu:i-j} always dominate, and the ${\mathcal O}(\varepsilon)$ corrections can be neglected. Note also, that if instead of \cref{eq:Lagr:Ui-j} we consider the $U(1)'_{\alpha-\beta}$ Lagrangian above the electroweak scale, the kinetic mixing is between hypercharge and $Z'$. 
After $SU(2)_L \times U(1)_Y \times U(1)' \to U(1)_{\text{EM}}$ breaking, there will now also be a mixing between ${Z}'$ and the SM $Z$. This gives a correction to $Z'$ mass, shifting  
$m_{Z'}=g'v' Q_\Phi$ to  $m_{{Z}'} (1 + \mathcal{O}(\varepsilon^2))$, and also induces tree-level couplings of $Z'$ to the SM Higgs $H$.

An important difference with the $U(1)'$ model (Model~I in \cref{sec:Model:I:kin:mix}), is that the LFV interactions can arise already at dimension-5
\begin{equation}\label{eq:LFV_lag_2}
    -\mathcal{L}_{\text{LFV}, \text{II}} = \frac{\tilde{k}_{ij}}{\Lambda} \Phi \bar{L}_i H \lepton_j + \frac{\tilde{k}'_{ji}}{\Lambda}\Phi^* \bar{L}_j H \lepton_i + \text{h.c.},
\end{equation}
if the $U(1)'_{\alpha-\beta}$ charge of $\Phi$ is $Q_\Phi=[L_i]_{\alpha-\beta}-[L_j]_{\alpha-\beta}$, where  $[L_i]_{\alpha-\beta}=[\ell_i]_{\alpha-\beta}$ is the charge of $i$--th generation SM lepton  under the $U(1)'_{\alpha-\beta}$.
 Focusing on $U(1)_{\mu-\tau}$, as a benchmark example, LFV dimension-5 operators are allowed by gauge invariance for
\begin{equation}
Q_\Phi = 
\begin{cases}
+2: & \tilde k_{\mu\tau}, \tilde k_{\tau\mu}' \ne 0, \\
+1: & \tilde k_{e\tau}, \tilde k_{\mu e}, \tilde k_{\tau e}', \tilde k_{e \mu}' \ne 0,  
\end{cases}
\end{equation}
where we indicated which matrix elements in $\tilde k$, $\tilde k'$ can be nonzero in each of the two cases. That is, in the $Q_\Phi=+2$ case only $\tau\to \mu$ FCNCs are allowed at dimension 5 level, while, in contrast, for $Q_\Phi=+1$ both $\tau\to e$ and $\mu \to e$ transitions are allowed, and $\tau \to \mu$ is not.

For the two numerical benchmarks we take
\begin{align}
\label{eq:benchmark:Lmu-Ltau:+2}
Q_\Phi=+2\text{~benchm.}&:  \tilde k_{\mu\tau}=\tilde k_{\tau\mu}'\equiv k_{\mu\tau},
\\
\label{eq:benchmark:Lmu-Ltau:+1}
Q_\Phi=+1\text{~benchm.}&:  \tilde k_{e\tau}=\tilde k_{\tau e}'\equiv k_{e\tau}, \,\, \tilde k_{\mu e}=\tilde k_{e \mu}' =0,
\end{align}
with $k_{\mu\tau}$ and $k_{e\tau}$ real (the motivation is simply a reduction in the number of parameters --- our results do not depend heavily on this simplification). Writing $\Phi=(v'+\phi)/\sqrt 2$, the dark scalar $\phi$ couples to the SM charged leptons, after the electroweak symmetry and the $U(1)'$ symmetry breaking. For the two benchmarks this interaction term is given by 
\beq
\label{eq:phi:couplings:Li-Lj}
 -\mathcal{L}_{\text{LFV}, \text{II}} \supset \frac{v}{2 \Lambda} k_{ij} \phi \bar{\ell}_{iL} \lepton_{jR} +  \text{h.c.},
\eeq
where  $k = \tilde{k} + \tilde{k}'$ is a symmetric $3\times 3$ matrix with as the only nonzero entries either the $k_{\mu\tau}=k_{\tau \mu}$ ($Q^{\mu\tau}_\Phi=+2$ benchmark) or $k_{e\tau}=k_{\tau e}$ ($Q^{\mu\tau}_\Phi=+1$ benchmark).

The dimension-5 interactions in \cref{eq:LFV_lag_2} also contribute to the charged lepton mass matrix, in addition to the usual SM Yukawa contribution, cf. \cref{eq:SM:Yukawa}. After mass diagonalization we thus have
\begin{equation}
\label{eq:rotation:Lmu:Ltau}
    \text{diag}(m_e, m_\mu, m_\tau) = V_L^\dagger\left(y + k\frac{v'}{\sqrt{2}\Lambda}\right)V_R \frac{v}{\sqrt{2}},
\end{equation}
with $V_{L,R}$ the two mass-matrix-diagonalizing unitary matrices. This is similar to the charged lepton mass matrix in dark photon model (Model I, \cref{eq:Model:I:mass:matrix}), except that now the corrections arise already at ${\mathcal O}(v'/\Lambda)$. Note also, that the $L_\mu-L_\tau$ charges are defined in the basis in which $y$ is diagonal, and allow for dimension-4 Yukawa's.

The $Z'$ couplings to the SM charged leptons are lepton flavor violating, which is in contrast to the dark photon model in Sec. \ref{sec:Model:I:kin:mix}.
The reason is that  in general the leptonic charge matrices $Q_{L,R}^{\mu - \tau}=\diag(0,1,-1)$ in \cref{eq:Jmu:i-j} do not commute with $V_{L,R}$, and thus induce flavor off-diagonal couplings to the $Z'$.
After transforming to the mass basis, the flavor structure of the $Z'$ couplings is encoded in
\begin{equation}
    V_L^\dagger Q_{\mu - \tau} V_L, \quad V^\dagger_R Q_{\mu - \tau} V_R.
\end{equation}
In the limit $k v' / 2 \Lambda \ll y_\tau$, the mixing matrices can then be parameterized as $V_{L,R} \simeq \mathbf{1} + \Theta_{L,R}$ with $\Theta^\dagger_{L,R} = - \Theta_{L,R}$, such that
\begin{equation}\label{eq:charge_in_mass_basis}
    V_{L,R}^\dagger Q_{\mu - \tau} V_{L,R} \simeq Q_{\mu - \tau} + [Q_{\mu - \tau},\Theta_{L,R}].
\end{equation}
For the two benchmarks this then gives for the nonzero entries in $\Theta_{L,R}$, 
\begin{equation}
\label{eq:Theta:i-j}
    (\Theta_{L,R})_{\ell3} =- (\Theta_{L,R})_{3\ell} \simeq  \frac{k_{\ell \tau}}{m_\tau}\frac{v' v}{2 \Lambda}, \quad \ell=e,\mu,
\end{equation}
where we neglected contributions suppressed by powers of $m_\ell/m_\tau$. The lepton current coupling to the $Z'$,  \cref{eq:Lagr:Ui-j}, is thus for the two benchmarks in \cref{eq:benchmark:Lmu-Ltau:+2} given by
\begin{equation}
\begin{aligned}
\label{eq:Jmu:mu-tau}
    J^\mu_{\mu-\tau} &= \bar{\mu} \gamma^\mu \mu - \bar{\tau} \gamma^\mu \tau + \bar{\nu}_\mu \gamma^\mu P_L \nu_\mu  - \bar{\nu}_\tau \gamma^\mu P_L \nu_\tau\\
    &+\frac{v' v}{2m_\tau \Lambda} \begin{cases}
    2 k_{\mu\tau}\ \bar{\mu} \gamma^\mu \tau~,  & Q_\Phi=2 \\
    k_{e\tau}\  \bar{e}\gamma^\mu\tau~,  & Q_\Phi=1
    \end{cases} ~+ {\rm h.c.},
\end{aligned}
\end{equation}
where we neglect corrections of ${\mathcal O}(\Theta_{L,R}^2)$.

The flavor violating couplings of $\phi$ to leptons are given by \cref{eq:phi:couplings:Li-Lj}. Rotating to the mass basis gives only suppressed corrections, which can be neglected to the precision we work. That is, in the notation of \cref{eq:generalL}, we have
\begin{equation}\begin{aligned}
    &c_{ij} = (V_L^\dagger k V_R)_{ij}\frac{v}{2\Lambda} \simeq k_{ij}\frac{v}{2\Lambda} + \mathcal{O}\left(\frac{k^2_{ij}}{m_j}\frac{v'v^2}{4\Lambda^2}\right),
\end{aligned}\end{equation}
and
\beq
\mu_V= 2 g'Q_\Phi m_{Z'}, \quad g_V = g'.
\eeq

\begin{figure*}[t!]
    \centering
    \includegraphics[width=0.85\textwidth]{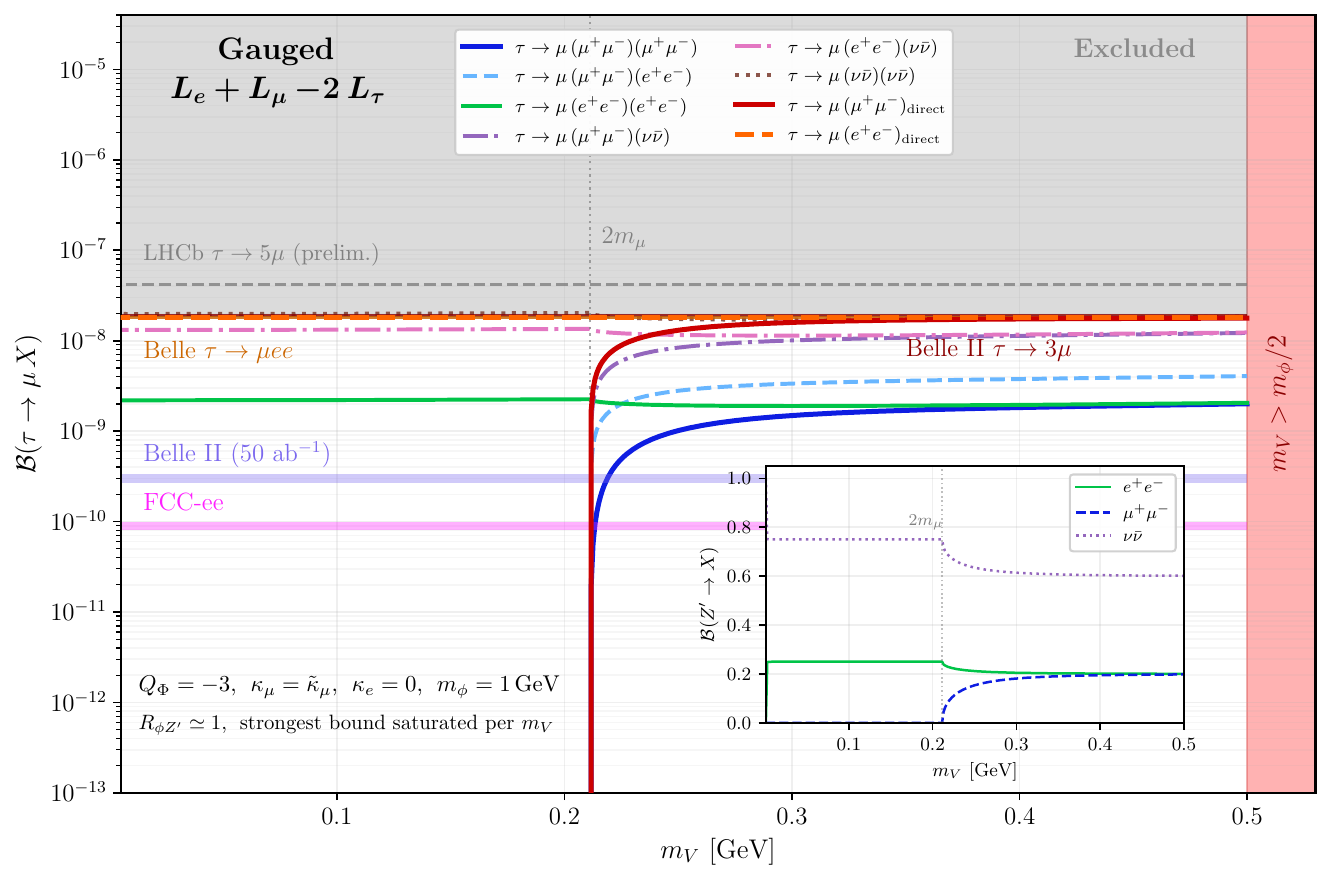}
    \caption{Same as \cref{fig:model1_BRs} but for the gauged $L_e + L_\mu - 2L_\tau$ model (Model III) with $m_\phi = 1$ GeV. In this model the $Z'$ carries lepton-flavor-violating couplings ($R_{\phi Z'}\simeq 1$), giving rise to both five-body cascade decays ($\tau\to\mu,\phi\to\mu,Z'Z'$) and direct three-body decays ($\tau\to\mu,Z'$). At each $m_V$ the LFV coupling is fixed by saturating the strongest existing bound among the Belle $\tau\to\mu e^+e^-$ (orange line, $1.8\times10^{-8}$), Belle II $\tau\to 3\mu$ limit (dark red line, $1.9\times10^{-8}$), and the preliminary LHCb $\tau\to 5\mu$ estimate of sensitivity (black line, $4.2\times10^{-8}$). 
    The gray shaded region above the Belle and Belle II lines is excluded. Because the $Z'$ decays to $e^+e^-$, $\mu^+\mu^-$, and $\nu\bar\nu$ with approximate branching ratios $0.2:0.2:0.6$, channels with missing energy dominate the cascade rate, while the purely leptonic five-body modes ($5\mu$, $\mu,4e$, $3\mu,2e$) account for $\sim 16\%$. \textit{Inset:} $Z'$ branching ratios as a function of $m_V$, showing the $\mu^+\mu^-$ threshold at $2m_\mu$. 
    }
    \label{fig:model3_BRs}
\end{figure*}

Note that in this model both $Z'$ and $\phi$ have LFV couplings, in contrast to the dark photon model in \cref{sec:Model:I:kin:mix}, where only the $\phi$ had LFV interactions.
The FCNC tau decays are thus both due to $\tau \to \ell \phi$ and $\tau \to \ell Z'$. 
In the $m_\phi, m_{Z'} \ll m_\tau$ limit, the branching ratios for the two channels are comparable
\begin{equation}
\label{eq:RphiZ'}
    R_{\phi Z'}^\ell\equiv \frac{\Gamma(\tau \to \phi \ell)}{\Gamma(\tau \to Z'\ell)}\simeq 1,
\end{equation}
as expected from the Goldstone boson equivalence theorem.
Therefore, in this model the $\tau \to 3 \mu$ and $\tau \to e2 \mu$ signatures, which one obtains from the $\tau \to \ell Z', Z'\to 2\mu$ decay cascades for the $Q^{\mu\tau}_\Phi = +2$ and $Q^{\mu\tau}_\Phi = +1$ benchmarks, respectively, compete with the $\tau \to 5 \mu$ and $\tau \to e4 \mu$ signatures  from the $\tau\to \ell \phi, \phi \to 2 Z', Z'\to 2\mu$ cascades. Since $\mathcal{B}(Z'\to 2\mu): \mathcal{B}(Z'\to 2\nu)\simeq 0.5:0.5$, we have
\beq
\frac{\Gamma(\tau \to \ell +4\mu)}{\Gamma(\tau \to  \ell +2\mu +2\nu ) }\simeq 0.5, \quad \ell=e,\mu,
\eeq
in both benchmarks. For other $L_\alpha-L_\beta$ choices one can obtain other multi-lepton signatures. For instance, for $U(1)'$ that gauges $L_e-L_\tau$ one would instead have $\tau \to \ell +2e$ and $\tau \to \ell +4e$ multibody decays, with $\ell =e,\mu$, for the choices of charge assignments for $\Phi$. 

Throughout this discussion we worked in the limit where kinetic mixing is negligible, i.e., $\varepsilon\to 0$ in \cref{eq:Lagr:Ui-j}. Consequently there were no couplings of $Z'$ to electrons and quarks. This is unlikely to be the case, however. That is, even if one forbids tree level couplings of $Z'$ to electrons and quarks, they will be induced at one-loop due to 
loop-induced kinetic mixing between hypercharge and $Z'$, and thus couplings of $Z'$ to the electromagnetic current. The effective mixing parameter depends on momentum $q$ flowing through the electromagnetic current, and is for $U(1)'$ that gauges $L_\mu-L_\tau$ given by~\cite{Amaral:2021rzw}
\begin{equation}
    \begin{aligned}
    \varepsilon_{\mu \tau}(q^2) &=  -\frac{e g'}{2\pi^2} \int_0^1dx \, x(x-1)\log \left( \frac{m_\mu^2 + q^2 x (x-1)}{m_\tau^2 + q^2 x(x-1)}\right).
    \end{aligned}
\end{equation}
For $Z'$ decays, $q^2=m_{Z'}^2$, so that the loop induced mixing is a constant. Assuming that the tree level $\varepsilon$ kinetic mixing parameter vanishes, the bounds on $L_\mu-L_\tau$ gauge coupling is $g'\lesssim \text{few} \times 10^{-4}$\,--\,$10^{-3}$ in the mass range of interest, $m_{A'}=0.1$\,--\,$1\,$GeV \cite{Ilten:2018crw,Greljo:2021npi}. It is, however, also possible that the sum of one loop and tree level contributions to the mixing, $\varepsilon_{\mu \tau}+\varepsilon$, cancel against each, in which case these bounds would get relaxed.

The LFV $Z'$ interactions in  \cref{eq:Jmu:mu-tau} induce $\tau \to \ell \gamma$ transition at one loop, due to a vertex correction with $Z'$ exchange.
The branching ratio $\mathcal{B}(\tau \to \ell \gamma)$ is suppressed by $\alpha' \alpha_\mathrm{EM}$ relative to $\mathcal{B}(\tau \to 3 \ell)$ and $\mathcal{B}(\tau \to 5 \ell)$ from dark cascades.
Since the current bounds on $\tau \to 3\ell$ and $\tau \to \ell \gamma$ are both of order $\mathcal{O}(10^{-8})$,
the bounds on $\tau \to 3\ell$ at present dominate the constraints on this model.

\subsection{Model III: gauged \texorpdfstring{$L_e +L_\mu - 2 L_\tau$}{\alpha L_X + \beta L_y + \gamma L_Z}}
\label{sec:model:III:gauge:all}

If the SM particle content is extended by $n_R$ right-handed neutrinos $N_i$, the linear combination of lepton numbers 
$U(1)'=a(L_\alpha - L_\beta) + b(L_\beta - L_\gamma)$ 
is anomaly free and can be gauged, if $a$ and $b$ satisfy $3 a b(a-b) = \sum_{i=1}^{n_R} Q_{N_i}^3$, with $Q_{N_i}$ the $U(1)'$ charges of right-handed neutrinos~\cite{Ballett:2019xoj,Costa:2019zzy,Allanach:2019gwp,Allanach:2020zna}.
Taking $a=1$, $b=0$ gives the models discussed in \cref{sec:Model:II:gauged:LX-LY}, which do not require right-handed neutrinos
for the anomaly cancellation.
Here, let us take $a=-b=1$, with  $\alpha=e$, $\beta=\tau$, and $\gamma=\mu$, so that 
\begin{equation}
\label{eq:Le_Lmu_2Ltau}
 U(1)'=L_e + L_\mu -2L_\tau.
\end{equation}
This is an anomaly free choice, if there are $n_R=3$ right-handed neutrinos with charges $Q_{N_1} = 1$, $Q_{N_2} = 1$, and $Q_{N_3} = -2$.
In general, we can introduce three right-handed neutrinos with $Q_{N_1} = a$, $Q_{N_2} = -b$, and $Q_{N_3} = b-a$ to cancel the anomaly
for any choice of $a$ and $b$.

\begin{figure*}[t!]
    \centering
    \includegraphics[width=0.99\textwidth]{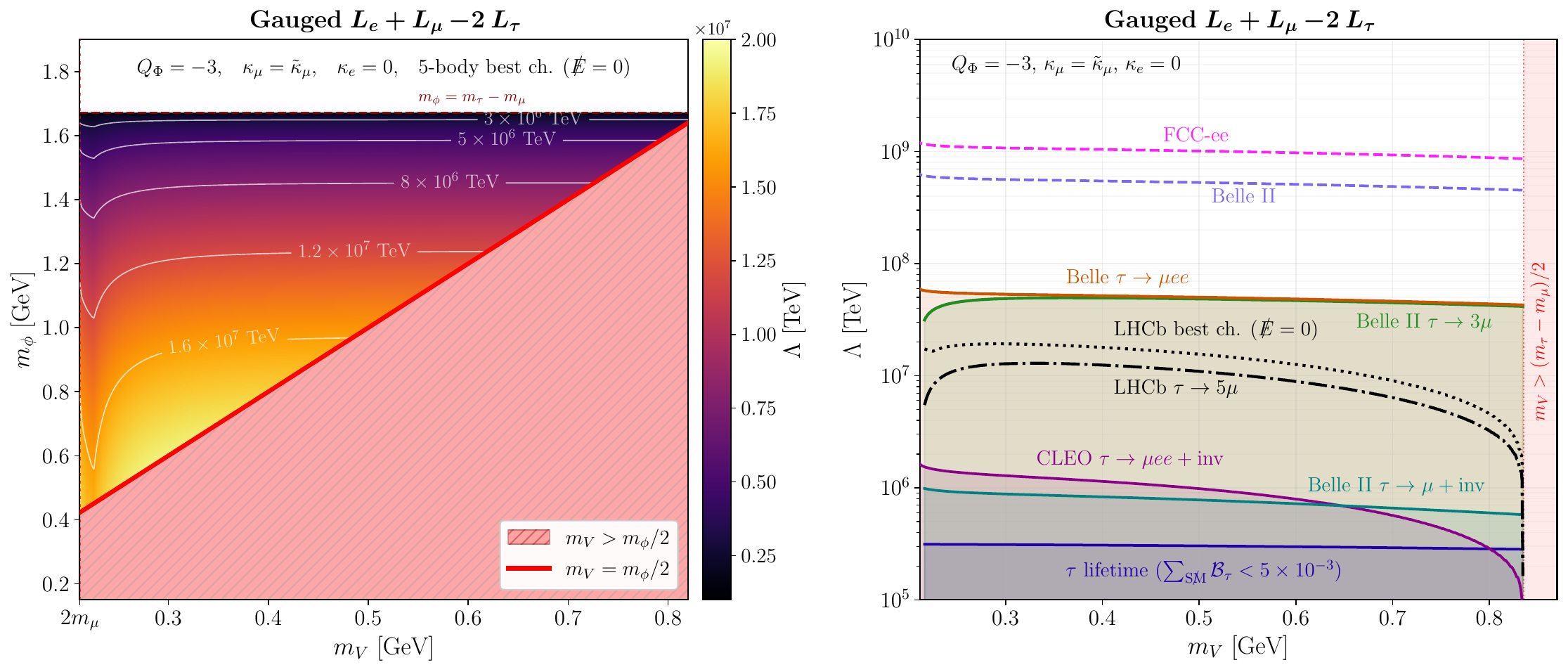}
    \caption{Same as \cref{fig:model1_Lambda} but for the gauged $L_e + L_\mu - 2L_\tau$ model (Model III) with dimension-5 LFV operators. 
    \textit{Left:} The heatmap shows the best fully visible five-body channel ($\slashed{E}=0$, i.e., excluding cascade final states with neutrinos) at each point, assuming comparable experimental sensitivity across channels.
    \textit{Right:} In addition to the $\tau$ lifetime (blue, $\sum \mathcal{B}_{\tau\to\slashed{\mathrm{SM}}} < 5\times10^{-3}$, maximized over all kinematically allowed $m_\phi$), the shaded regions show existing constraints from the direct three-body decays $\tau\to 3\mu$ (Belle II, green) and $\tau\to\mu, e^+e^-$ (Belle, orange), as well as the semi-invisible cascade channel $\tau\to\mu, e^+e^- + \mathrm{inv}$ (CLEO, purple) and $\tau\to\mu + \mathrm{inv}$ (Belle II, teal). All other curves are maximized over $m_\phi > 2m_V$ (on-shell cascade regime). 
    For this model $R_{\phi Z'}\simeq 1$, so the direct $\tau\to\ell, Z'$ channels compete with the five-body cascades; the Belle~II and FCC-ee projections reflect the best sensitivity across both topologies. In both panels $Q_\Phi = -3$, $\kappa_\mu = \tilde\kappa_\mu$, $\kappa_e = 0$.
    }\label{fig:model3_Lambda}
\end{figure*}

The $U(1)'$ gauge group is spontaneously broken once the scalar field $\Phi$ charged under it, obtains a VEV. If the $U(1)'$ charge of $\Phi$ is $Q_{\Phi} = -3$,
then the dimension-5 LFV interactions necessarily involve the third generation leptons and are given by
\beq
\begin{split}
\label{eq:ModelIV:Lagr}
    - \mathcal{L} &\supset \frac{\kappa_e}{\Lambda} \, \Phi \bar{L}_\tau H e_R + \frac{\tilde \kappa_e}{\Lambda} \, \Phi^* \bar{L}_e H \tau_R 
    \\ 
    & \quad + \frac{\kappa_\mu}{\Lambda} \Phi \bar{L}_\tau H \mu_R + \frac{\tilde\kappa_\mu}{\Lambda} \Phi^* \bar{L}_\mu H \tau_R + \mathrm{h.c.}.
    \end{split}
\eeq
That is, for $Q_{\Phi} = -3$ charge assignment the dimension-5 LFV operators will lead to $\tau \to \mu \phi$ and/or $\tau\to e\phi$ decays, while the $\mu\to e \phi$ transitions will be further suppressed even if kinematically allowed.   
When $H$ and $\Phi$ obtain VEVs, the above Yukawa like interactions in \cref{eq:ModelIV:Lagr} contribute to the charged lepton mass matrix. For simplicity let us take as a benchmark (similarly to the benchmark for the $L_\mu-L_\tau$ model in the previous subsection) 
\beq
\kappa_e=\tilde \kappa_e, \qquad \kappa_\mu=\tilde \kappa_\mu, 
\eeq
and assume that both $\kappa_e$ and $\kappa_\mu$  are defined in the basis where the $e$\,--\,$\mu$ block of the SM Yukawa matrix $y_{ij}$ has been diagonalized.
Rotating to the mass basis as in \cref{eq:rotation:Lmu:Ltau}, we have $V_{L,R} \simeq \mathbf{1} + \Theta_{L,R}$, where, neglected contributions suppressed by powers of $m_\ell/m_\tau$,
\begin{equation}
\label{eq:Theta:e+mu-2tau}
    (\Theta_{L,R})_{\ell3} =- (\Theta_{L,R})_{3\ell} \simeq  \frac{\kappa_{\ell}}{m_\tau}\frac{v' v}{2\Lambda}, \quad \ell=e,\mu.
\end{equation}
The $L_e+L_\mu-2L_\tau$ lepton current coupling to the $Z'$, ${\cal L}_{\rm int}\supset g' Z_\mu' J^\mu_{e+\mu-2\tau}$, is 
 in the mass basis given by
\begin{equation}
\begin{aligned}
\label{eq:Jmu:e+mu-2tau}
    J^\mu_{e+\mu-2\tau} &= \bar{e} \gamma^\mu e+\bar{\mu} \gamma^\mu \mu - 2 \bar{\tau} \gamma^\mu \tau 
    \\
    &+ \bar{\nu}_e \gamma^\mu P_L \nu_e+ \bar{\nu}_\mu \gamma^\mu P_L \nu_\mu  - 2 \bar{\nu}_\tau \gamma^\mu P_L \nu_\tau
    \\
    &+ 3\frac{\kappa_{\mu}}{m_\tau}\frac{v' v}{2\Lambda} \bar{\mu} \gamma^\mu \tau + 3 \frac{\kappa_{e}}{m_\tau} \frac{v' v}{2\Lambda} \bar{e}\gamma^\mu\tau +{\rm h.c.},
\end{aligned}
\end{equation}
where we neglect corrections of ${\mathcal O}(\Theta_{L,R}^2)$. As expected for a model where the $\Phi$ coupling is a dimension-5 operator
\begin{equation}
\label{eq:RphiZ':Le:Lmu:Ltau}
    R_{\phi Z'}^\ell= \frac{\Gamma(\tau \to \phi \ell)}{\Gamma(\tau \to Z'\ell)}\simeq 1.
\end{equation}

The decays of the $U(1)'$ gauge boson are, up to phase space considerations, the same for electrons and muons, namely $\Gamma_{Z' \to ee} \simeq \Gamma_{Z'\to \mu \mu}$. Assuming that the sterile neutrinos are heavier, we thus have $\mathcal{B}(Z' \to ee)\simeq \mathcal{B}(Z' \to \mu \mu )\simeq \mathcal{B}(Z' \to \nu \nu)=0.2:0.2:0.6$. The $\tau \to \ell \phi$, $\phi \to 2Z'$ cascade decays will thus lead to a number of multi-lepton decays. The decays channels with odd number of electrons + positrons in the final state arise when $\kappa_e, \tilde \kappa_e\ne 0$, with branching ratios that are controlled by the $Z'$ decay branching ratios, 
\begin{equation}\label{eq:odd_e_numbers}
\begin{split}
    &\mathcal{B}(\tau \to 5 e):  \mathcal{B}(\tau \to e 4\mu):  \mathcal{B}(\tau \to 3e 2\mu) :
    \\
     &\mathcal{B}(\tau \to 3e 2\nu) : \mathcal{B}(\tau \to e 2\mu 2\nu):\mathcal{B}(\tau \to e 4\nu)
     \\
     &\simeq 0.04:0.04:0.08:0.24:0.24:0.36.
 \end{split}
\end{equation}
Similarly, if  $\kappa_\mu, \tilde \kappa_\mu\ne 0$, one obtains final states with odd number of muons. 
\begin{equation}\label{eq:odd_muon_numbers}
\begin{split}
    &\mathcal{B}(\tau \to 5 \mu):  \mathcal{B}(\tau \to \mu 4e ):  \mathcal{B}(\tau \to 3\mu 2 e) :
    \\
     &\mathcal{B}(\tau \to 3\mu 2\nu) : \mathcal{B}(\tau \to \mu 2 e 2\nu):\mathcal{B}(\tau \to \mu 4\nu)
     \\
     &\simeq 0.04:0.04:0.08:0.24:0.24:0.36.
 \end{split}
\end{equation}
The relative sizes between the two sets of branching ratios depend on the values of $\kappa_i$ and are, ignoring phase space effects, given by
\beq
\frac{\mathcal{B}(\tau \to \text{odd}\,e)}{\mathcal{B}(\tau \to \text{odd} \,\mu)}\simeq\frac{|\kappa_e|^2+| \tilde \kappa_e|^2}{|\kappa_\mu|^2+| \tilde \kappa_\mu|^2}.
\eeq

The predicted branching ratios as a function of $m_V$ are shown in \cref{fig:model3_BRs}, and the resulting bounds on $\Lambda$ are shown in \cref{fig:model3_Lambda}.
The projected Belle~II and FCC-ee sensitivities in \cref{fig:model3_Lambda} are obtained by rescaling current $\tau \to 3\mu$ and $\tau \to \mu\, e^+e^-$ upper limits by the expected increase in tau statistics (see \cref{sec:exp:prospects}).
As a benchmark we set $\kappa_\mu = \tilde\kappa_\mu \neq 0$ with $\kappa_e = \tilde\kappa_e = 0$, which is the minimal choice that induces exotic tau decays with odd muon number while keeping the electron channels in \cref{eq:odd_e_numbers} closed.
Note that unlike Model~I, no separate flavor-diagonal scalar coupling is needed for the $\tau \to 3\mu$ constraint to apply: in this model $R_{\phi Z'}^\ell \simeq 1$, so the $\tau \to \mu Z'$, $Z' \to \mu^+\mu^-$ cascade contributes to $\tau \to 3\mu$ at the same order as the five-body channels, with the flavor-diagonal $Z'$ couplings fixed by the gauge structure.
A more general flavor structure with $\kappa_{e},\tilde{\kappa}_e \neq 0$ would open the additional odd-electron channels in \cref{eq:odd_e_numbers} but does not change the qualitative conclusions. 
Among the above exotic decay channels, the $\tau \to \ell 4\nu$ decays have the largest branching ratios. However, they also face a very large SM background from tree level leptonic tau decays. The main sensitivity is thus expected to come from tau decays to five charged leptons. The $\tau \to \ell 2 \ell' 2\nu$ decays also face SM backgrounds, however, one can still search for exotic decays by searching for a bump in the $2\ell'$ invariant mass.

The $L_e+L_\mu-2 L_\tau$ charge assignment is not the only one that can lead to $\tau\to \ell \phi$ FCNCs with no (or with only highly suppressed) $\mu \to e \phi$ transitions. For instance, 
$L_e+2 L_\tau -3 L_\mu$, with $Q_\Phi=5$ would give a dimension-5 interaction Lagrangian as in \eqref{eq:ModelIV:Lagr}, but now with only $\tau\to \mu \phi $ transitions allowed by gauge invariance. Similarly, $2 L_e+ L_\tau -3 L_\mu$, with $Q_\Phi=5$ would only give rise to $\tau\to e \phi $ decays.

We note that it is possible to decouple the $\tau \to \ell \phi$ branching ratio from that of $\tau \to \ell Z'$ in variants of this model.
In particular, we may consider instead gauging $L_\mu + L_\tau - 2L_e$ and the dimension-6 operators 
\begin{equation}\label{eq:phi_squared_L}
    -\mathcal{L} \supset \frac{\kappa'}{\Lambda^2} |\Phi|^2 \bar L_\tau H \mu_R + \frac{\tilde\kappa'}{\Lambda^2} |\Phi|^2 \bar L_\mu H \tau_R +{\rm h.c.}.
\end{equation}
The only constraint on the charge of $\Phi$ is that they are such that the dimension-5 operators are forbidden.
Note that at tree-level, the electron sector remains unaffected by the new scalar, and so no $e$\,--\,$\mu$ or $e$\,--\,$\tau$ LFV is generated.
In this scenario, the diagonalization of the lepton mass matrix is straightforward and commutes with the charge matrix $Q_{\mu+\tau-2e}$,
since the gauge invariance forbids SM Yukawa couplings between $e$\,--\,$\mu$ and $e$\,--\,$\tau$, resulting in the rotation only in the $\mu$\,--\,$\tau$ sector.
Therefore, by virtue of \cref{eq:charge_in_mass_basis}, we have 
\begin{align}
    J^\mu_{\mu+\tau-2e} &= \bar\tau\gamma^\mu\tau + \bar\mu\gamma^\mu \mu - 2\bar e \gamma^\mu e
    \\\nonumber
    &+ \bar{\nu}_e \gamma^\mu P_L \nu_e+ \bar{\nu}_\mu \gamma^\mu P_L \nu_\mu  - 2 \bar{\nu}_\tau \gamma^\mu P_L \nu_\tau.
\end{align}
As expected, no LFV interactions appear involving the $Z'$.
The scalar interactions in the mass basis are
\begin{equation}
    \mathcal{L} \supset \frac{\kappa' v'v}{\sqrt2 \Lambda^2} \phi \left( \bar\tau \mu + \bar\mu \tau\right),
\end{equation}
where we omit $\phi^2$ terms that would also appear with the operators of \cref{eq:phi_squared_L} as well as the LFC terms that are of higher order.

\subsection{Model IV: chiral model}
\label{sec:Model:IV:chiral:model}

\begin{table}[t]
   \centering
   \begin{tabular}{l  c c c c c c l} 
  \multicolumn{8}{c}{Model IV: benchmark chiral models}\\
  \hline\hline
   Model & $c_{R}$ & $b_{R}$ & $\mu_{R}$ & $e_{R}$ & $\nu_R$  & $\Phi$ &Channel  \\
    \hline
  Chiral $U(1)_\mu$ & 1         &   -1           &    -1          &     0     &     1         &  1 &  $\tau \to 5\mu$  \\
  Chiral  $U(1)_e$  & 1         &   -1           &    0          &     -1     &     1         &  1 & $\tau \to 5e$ \\
       \hline
       \hline
    \end{tabular}
   \caption{Charge assignment in the two chiral $U(1)'$ benchmark models, that either result in $\tau \to 5\mu$ or $\tau \to 5e$ signatures, see  \cref{sec:Model:IV:chiral:model} (Model IV) for details.
   The SM fields not shown are neutral under the $U(1)'$.}
   \label{tab:chiralmodel}
\end{table}

An interesting possibility is that the SM fermions, both quarks and leptons, are charged under the additional $U(1)'$ -- we will focus on the cases where only the right-handed SM fermions carry a $U(1)'$ charge. As in previous cases, we assume that $U(1)'$ is spontaneously broken once scalar  $\Phi$ charged under it obtains a vev, so that $\Phi=(v'+\phi)/\sqrt{2}$. The light scalar $\phi$ interacts with the SM fermions via dimension 5 operators (summation over repeated indices is assumed)
\be\label{eq:model_IV_Lagrangian}
\begin{split}
\mathcal{L}_{\text{IV}} \supset &\frac{a_{ij}}{\Lambda} \Phi \bar{L}_i H \ell_j + \frac{b_{ij}}{\Lambda} \Phi \bar{Q}_i H d_{R_j} + \frac{b'_{ij}}{\Lambda} \Phi \bar{Q}_i \tilde H u_{R_j}
\\
+ &
\frac{\tilde a_{ij}}{\Lambda} \Phi^* \bar{L}_i H \ell_j + \frac{\tilde b_{ij}}{\Lambda} \Phi^* \bar{Q}_i H d_{R_j} + \frac{\tilde b'_{ij}}{\Lambda} \Phi^* \bar{Q}_i \tilde H u_{R_j}
\\
+&\text{ h.c.}.
\end{split}
\ee
Which of the Yukawa-like couplings $a_{ij}$, $\tilde a_{ij}$, $b^{(\prime)}_{ij}$ or $\tilde b^{(\prime)}_{ij}$ are allowed by $U(1)'$ gauge invariance depends on the concrete model, i.e., on the charge assignments of the SM fermions and $\Phi$. 
While there are many generation-dependent chiral assignment of $U(1)'$ gauge charges that are anomaly free once sterile neutrinos are included~\cite{Allanach:2018vjg}, we focus on a particular subset: those in which $Z'$ couples to only one generation of right-handed charged leptons, and only to heavy quarks, $c,b,$ or $t$. For concreteness we choose two benchmark models ``Chiral $U(1)_\mu$'' and ``Chiral $U(1)_e$'', with charge assignments shown in \cref{tab:chiralmodel}. In both $Z'$ couples to $c_R, b_R$, $\Phi$ and a single sterile neutrino $\nu_R$ (necessary to cancel anomalies), while in ``Chiral $U(1)_\mu$'' $Z'$ also couples to muons, and in ``Chiral $U(1)_e$'' to electrons. In the two benchmark models thus
\begin{align}
\text{Chiral $U(1)_\mu$}: & 
\begin{cases}
a_{e\mu}, a_{\tau\mu}\ne 0,
\\  
\tilde b'_{uc}, \tilde b'_{tc}\ne 0,
 \\
  b_{db}, b_{sb}\ne 0,
  \end{cases}
  \\
  \text{Chiral $U(1)_e$}: & 
\begin{cases}
a_{\mu e}, a_{\tau e}\ne 0,
\\  
\tilde b'_{uc}, \tilde b'_{tc}\ne 0,
 \\
  b_{db}, b_{sb}\ne 0,
  \end{cases}
\end{align}

In both cases, the $\mu\to e \phi$ transition is in principle possible. However,  we are primarily interested in the part of parameter space where this decay is not kinematically allowed, $m_\phi> m_\mu-m_e$, as is the case, e.g., for our numerical benchmark \cref{eq:benchmark_onshell}. The main signal for the two benchmark models will thus be $\tau \to 5\ell$, i.e., for Chiral $U(1)_\mu$ this is $\tau \to 5\mu$  from $\tau \to \mu \phi, \phi \to 2 Z', Z'\to 2\mu$ decay cascade, and similarly the main exotic signal is $\tau \to 5e$ for Chiral $U(1)_e$. The decay width for $\tau\to \ell \phi$ is given in \cref{eq:partialwidthtueatophimu} with $c_{\tau\ell}= a_{\tau\ell}v/2 \Lambda$, giving in the $m_\phi, m_\ell \ll m_\tau$ limit,
\beq
\mathcal{B}(\tau\to \ell \phi)\simeq 3\cdot 10^{-6} \biggr( \frac{10^{10}\,\text{GeV}}{\Lambda/a_{\tau\ell} }\biggr)^2.
\eeq
Note that in the two models $Z'$ decays almost exclusively into $Z'\to 2\mu$ and $Z'\to 2e$, respectively, for $Z'$ masses below open charm threshold, i.e., in the kinematical regime we are interested in (if $\nu_R$ are light enough, then also $Z'\to$inv is possible).  As with previous models where the flavor violating coupling of $\phi$ is dimension 5 there are flavor violating vector couplings and
\begin{equation}\label{eq:RphiZchiral}
    R_{\phi Z'}^\ell= \frac{\Gamma(\tau \to \phi \ell)}{\Gamma(\tau \to Z'\ell)}\simeq 1~.
 \end{equation}

In both benchmark models also $c\to u\phi (Z'), t\to c \phi (Z')$ and $b\to s\phi (Z'), b\to d\phi (Z')$ transitions are possible. We focus on the decay widths for $P_1\to P_2\phi$ decays, induced by $c\to u\phi$ or $b\to s\phi, d\phi$, although the vector decays will have similar behavior due to \cref{eq:RphiZchiral}.  
The scalar decays are given by (for only $b_{ij}$ nonzero)
\beq
\Gamma(P_1\to P_2 \phi)=\frac{M_1}{16\pi}\biggr(\frac{v\, \text{Re}\,b_{ij}}{2\Lambda}\biggr)^2 f_0^2(m_\phi^2) \zeta_\phi,
\eeq
where $\zeta_\phi\simeq 1$ for $m_{P_2}, m_\phi \ll m_{P_1}$. Explicitly, for $q_1\to q_2 \phi$ transitions, it is given by
\beq
\zeta_\phi=\biggr(\frac{M_1^2-M_2^2}{m_{q_1}-m_{q_2}}\biggr)^2\frac{ \lambda^{1/2}(M_1^2, M_2^2,m_\phi^2)}{ M_1^4},
\eeq
with $m_{q_i}$ the mass of $q_i$ quark, $\lambda(M_1^2, M_2^2,m_\phi^2)= M_1^4+\cdots$ is the K\"allen function, $M_{1,2}$ are the masses of $P_{1,2}$ mesons, and $ f_0(q^2)$ is the $P_1\to P_2$ form factor relevant for scalar currents.

For simplicity we have also taken $b_{ij}^{(\prime)}$ to be real (if $b_{ij}^{(\prime)}$ are complex, in addition two body meson decays to vector mesons, $P\to V\phi$, are possible). Numerically, 
\begin{align}
\mathcal{B}(D^+\to \pi^+ \phi)&\simeq 3\cdot 10^{-6} \biggr( \frac{10^{10}\,\text{GeV}}{\Lambda/\text{Re}\,\tilde b'_{uc} }\biggr)^2,
\\
\mathcal{B}(B^+\to K^+ \phi)&\simeq 4\cdot 10^{-6} \biggr( \frac{10^{10}\,\text{GeV}}{\Lambda/\text{Re}\,b_{sb} }\biggr)^2,
\\
\mathcal{B}(B^+\to \pi^+ \phi)& \simeq 2\cdot 10^{-6} \biggr( \frac{10^{10}\,\text{GeV}}{\Lambda/\text{Re}\,b_{db} }\biggr)^2.
\end{align}
These decays lead to $P_1\to P_2 +4\mu$ and  $P_1\to P_2 +4 e$  exotic signatures via cascade decays, $P_1\to P_2 \phi, \phi\to 2 Z', Z'\to 2\ell$. Whether $\tau \to 5\ell$ or $P_1\to P_2 + 4\ell$ are the more important search channels, depends on the flavor structure in the UV, i.e., on the values of the $a_{ij}$ and $b_{ij}^{(\prime)}$ coefficients. The above models also in principle induce exotic top decays. However, since the SM top decay width is not suppressed, the exotic top decays are outside experimental reach, unless all other coefficients are extremely, suppressed, since $\mathcal{B}(t\to c \phi)\simeq {\mathcal O}(10^{-16} )$ for $\Lambda/b_{tc}\sim {\mathcal O}(10^{10}\,\text{GeV})$. Similarly, the tree level exchanges of $\phi$ induce corrections to $D$\,--\,$\bar D$, $B$\,--\,$\bar B$, $B_s$\,--\,$\bar B_s$ mixing amplitudes that are well below any current or future sensitivities.

\subsection{Model V: flavor protected scalar}
\label{sec:model:V:flavor:protected:scalar}

As the final example let us discuss the possibility that the SM field content is extended by a light flavor protected scalar, $\cS$. In this model, the  cascade $\tau \to \mu 3\cS$, $\cS\to 2\ell$ results in the $\tau \to \mu +6\ell$ decay as the main exotic signature.  The model is a modification of the flavor protected scalar model, which was introduced in Ref.~\cite{Greljo:2025ljr} as an example of a natural model that can lead to a $\mu \to 7 e$ exotic decay signature. 
At low energies the interactions between $\cS$, a SM singlet, and the SM fields include the following higher dimension operators, 
\begin{equation}\label{eq:opsModelI}
\mathcal{L} \supset \frac{\cC_{\cS}^{(3)}}{3!\Lambda^3}\,\bar L_{\mu} H \tau_{R}\, \cS^3 + \frac{\cC_{\mu\cS}}{\Lambda}\bar L_{\mu} H \mu_{R} \cS + \frac{\cC_{e\cS}}{\Lambda}\bar L_{e} H e_{R} \cS+ \text{h.c.}\,.
\end{equation}
For $\cS$ light enough, with a mass $m_\cS<(m_\tau-m_\mu)/3$, the $\cS^3$ term leads to $\tau \to \mu 3\cS$ decays. The second interaction facilitates $\cS\to 2\mu$ decays, so that the $\tau \to \mu 3\cS$, $\cS\to 2\mu$ cascade gives rise to $\tau \to 7\mu$ signature, assuming for simplicity $\cC_{\mu\cS}\gg \cC_{e\cS}$. If instead,  $\cC_{\mu\cS}\ll \cC_{e\cS}$, the main decay cascade is $\tau \to \mu 3\cS$, $\cS\to 2e$, leading to the $\tau \to \mu6e$ signature. Finally, if $\cC_{\mu\cS}$ and $\cC_{e\cS}$ are comparable, then  $\tau \to 7\mu$, $\tau\to 5\mu 2e$, $\tau\to 3\mu 4e$, and $\tau\to \mu 6e$ decay channels are all important. 

The fact that these high charged lepton multiplicity decays are the main exotic signatures is a result of a global $U(1)_\ell^3\equiv U(1)_e\times U(1)_\mu\times U(1)_\tau$ symmetry, broken by two spurions, $\varepsilon_\mu$ and $\varepsilon_\tau$. The decays $\tau \to \mu \cS$ and $\tau \to \mu 2\cS$ are in this case  parametrically suppressed (we assume that $\cal S$ does not obtain a vacuum expectation value). Under $U(1)_\ell^3$  the SM charged lepton fields carry the usual $L_\alpha$ charges 
while the light scalar has $[\cS]_{L_\tau}=\frac{1}{3}$. The SM flavor diagonal charged lepton Yukawa interactions are allowed by the $U(1)_\ell^3$ global symmetry, with the off-diagonal terms forbidden, thus defining the mass basis for the SM charged leptons.

The $U(1)_\ell^3$ global symmetry is not exact, but rather explicitly broken by two small spurions, with charges $[ \varepsilon_\tau ]_{L_\tau} = -\frac{1}{3} $ and $[ \varepsilon_\mu ]_{L_\mu}=-1$. The Wilson coefficients in \cref{eq:opsModelI} are therefore of order
\beq
\cC_{\cS}^{(3)}\sim {\mathcal O}(\varepsilon_\mu)\,, 
\qquad 
\cC_{\mu\cS}\sim \cC_{e\cS}\sim {\mathcal O}(\varepsilon_\tau)\,.
\eeq
That is, the $S\to 2\mu$ and $S\to 2e$ decay rates are suppressed by $\varepsilon_\tau^2$, while the $\tau \to \mu  3\cS$ decay rate is suppressed by $\varepsilon_\mu^2$.
In contrast, the operators that would give rise to $\tau\to \mu \cS$ and $\tau \to \mu  2\cS$ decays are further suppressed, and are of order,
\beq
\label{eq:suppressed:expected}
\bar L_\mu H \tau_R \cS^\dagger\sim {\mathcal O}(\varepsilon_\mu \varepsilon_\tau^2)\,, 
\qquad 
\bar L_\mu H \tau_R \cS^{\dagger 2}\sim {\mathcal O}(\varepsilon_\mu \varepsilon_\tau)\,,
\eeq
where the scaling refers to the parametric size of Wilson coefficients multiplying the operators. 
If $\varepsilon_\tau \ll m_\tau/(4\pi \Lambda)$, there is a hierarchy among the exotic decays, $\Gamma(\tau \to \mu \cS)\ll \Gamma(\tau \to \mu  2\cS) \ll \Gamma(\tau \to \mu 3\cS)$, so that the $\tau\to \mu+6\ell$ signature is the dominant one. Note that $\varepsilon_\mu$ enters in the same way into all three decay widths, and thus $\tau \to \mu 3\cS$ can dominate even for $\varepsilon_\mu$ that is large, $\varepsilon_\mu\sim {\mathcal O}(1)$. Furthermore, we have assumed that the $U(1)_e$ global symmetry is unbroken. If instead it is broken by a spurion $\varepsilon_e \gg \varepsilon_\mu$, with $[ \varepsilon_e ]_{L_e}=-1$, the above discussion changes trivially, by simply swapping $\mu\leftrightarrow e$ everywhere. In particular, the leading exotic signatures are now $\tau\to e +6\ell$. Note furthermore that $\mu\to e \cS$ would be allowed by the $U(1)_\ell^3$ is one has both a spurion that is charged under $U(1)_e$, $[\varepsilon_e]_{L_e}=-1$, as well as a spurion charged under $U(1)_\mu$, $[\varepsilon_\mu]_{L_\mu}=-1$.  However, the $\mu\to e n \cS$ decays are then suppressed both by $\epsilon_\mu$ as well as $\epsilon_e$ and powers of $\epsilon_\tau$, and can thus be very small (if $\cS$ is heavier than $m_\mu-m_e$ then they are also kinematically forbidden). 

The exotic decay width is 
\beq\label{eq:tau_to_mu_3S}
\Gamma (\tau \to \mu 3\cS) = \frac{m_\tau^5}{9216\pi}\biggr(\frac{v_\text{EW}}{\Lambda^3} \cC_\cS^{(3)}\biggr)^2 \biggr(\frac{1}{16\pi^2}\biggr)^2
~,
\eeq
where we have worked in the massless limit for the final state particles.  If the decay of the $\tau$ takes place instead near threshold the width scales as $\left( 1 - \xi \right)^{7/2}$ where
$\xi = 3m_{\cS}/(m_\tau - m_\mu)$  accounts for the phase-space suppression of the decay rate.
Ignoring this additional phase space factor, the branching ratio is
\beq
\mathcal{B} (\tau \to \mu 3\cS) \sim 7\cdot 10^{-10} \,\big(\cC_\cS^{(3)}\big)^2\biggr(\frac{1\, \text{TeV}}{\Lambda}\biggr)^6.
\eeq
Assuming that  $\cS\to 2e  $ dominates, i.e., taking $\cC_{e \cS}\gg \cC_{\mu \cS}$ for simplicity, so that we do not need to take into account the phase space dependence on final state masses, we have
\begin{equation}
\label{eq:ctauLcS}
    c\tau_\cS =109\,\mathrm{\mu m}
    \bigg( \frac{10^{-4}}{\cC_{e \cS}} \bigg)^2
    \bigg( \frac{\Lambda}{1\, \text{TeV}}\bigg)^2
    \bigg( \frac{300\,\text{MeV}}{m_\cS} \bigg)\,.
\end{equation}
In the region of parameter space we are interested,  the $\cS$ decays are thus  essentially prompt, though if $\cC_{e \cS}$ is significantly smaller than $10^{-4}$ the decays of $\cS$ will be displaced.

The above numerical benchmark estimates show that there is a large available parameter space, where $\tau\to \mu 3\cS$ would be the dominant exotic decay, for instance, $\cC_{e\cS}\sim \epsilon_\tau \sim 10^{-4}\ll m_\tau/(4\pi \Lambda)$ for $\Lambda\sim{\mathcal O}(\text{TeV})$. These rare tau decays would probe new physics mass scales in the TeV regime, and would thus be complementary to direct searches at the LHC (for a UV completion of the above model, involving a scalar and a vector-like lepton, see  Ref.~\cite{Greljo:2025ljr}, however, now with a modified flavor pattern).

\section{Constraints} \label{sec:constraints}
The models for multilepton tau decays that we introduced in the previous section face several existing experimental constraints. For models I to IV these are the constraints: (i) on LFV $\tau$\,--\,$\ell$\,--\,$\phi$ couplings and on LFV $\tau $\,--\,$\ell$\,--\,$V$ couplings; (ii) on flavor diagonal couplings of $\phi$ as well as  $\phi$\,--\,$V$\,--\,$V$ couplings; and (iii) on flavor diagonal couplings of $V$ to leptons and pions. For model V, we similarly have the constraints on flavor violating $\tau$\,--\,$\ell$\,--\,$3\cS$ couplings, and on flavor diagonal couplings of $\cS$. The constraints on flavor diagonal couplings of $\phi$, $V$ and $\cS$,  
are not specific to the LFV models introduced in \cref{sec:benchmark:models}, and have been extensively discussed in the literature. For instance, the general bounds on a light vector coupling to SM fermions, allowing for general values of flavor diagonal couplings can be found in \cite{Ilten:2018crw,Baruch:2022esd}. 
 Constraints on general flavor diagonal light scalars coupling to SM fermions can be found in \cite{AxionLimits}, while more specific constraints on lepton couplings for various leptophilic $\phi$ benchmarks, in the mass regime we are interested, can be found in \cite{Ema:2025bww} (note, however, that for the phenomenology discussed in the current manuscript the most relevant part of the parameter space is when the dominant decay mode of the light scalar $\phi$ is the decay $\phi\to 2V$). 

Below, we thus focus on the existing experimental constraints on the LFV coupling $c_{ij}\, \phi \,\bar{\ell}_{L,i} \ell_{R,j}$,  cf. \cref{eq:generalL}, which is the common component of the models I-IV presented in Sec. \ref{sec:benchmark:models}. We will also discuss constraints on $\tau$\,--\,$\ell$\,--\,$3\cS$ coupling LFV coupling present in model V. 

We show explicit results for Models~I and III in \cref{fig:model1_BRs,fig:model1_Lambda,fig:model3_BRs,fig:model3_Lambda}. 
Model~I is representative of scenarios in which the LFV coupling arises at dimension 6, so that $\tau \to \ell V$ is suppressed relative to $\tau \to \ell \phi$ and all final states are fully visible. 
Model~III is representative of the class of models (including Models~II and IV) in which the LFV coupling arises at dimension 5, implying $R_{\phi Z'}^\ell \simeq 1$ (see \cref{eq:RphiZ',eq:RphiZ':Le:Lmu:Ltau,eq:RphiZchiral}), so that the direct three-body decays $\tau \to \ell V$ compete with the five-body cascades. 
In Models~II and III, $V \to \nu\bar\nu$ decays additionally open missing energy channels that are absent in Models~I and IV.

The bounds on LFV couplings are both due to direct  and indirect constraints:

\paragraph{\bf Two body $\boldsymbol{\tau \to \lepton \phi_{\mathrm{inv}}}$ decays.} The searches for two-body $\tau\to \ell +$inv decays are relevant in two cases: (i) if $\phi$ is so weakly coupled that it decays outside the detector, or (ii) if $\phi$ decays invisibly.  The limit of effectively stable $\phi$ that escapes detection is relevant in all four models I to IV in Sec. \ref{sec:benchmark:models}. It occurs when 
the $\phi$\,--\,$V$\,--\,$V$ coupling $\mu_V$ in \cref{eq:generalL} is very small, $\mu_V\ll m_\phi$, 
so that a significant fraction of $\phi\to VV$ decays occurs outside the detector.  The possibility of invisibly decaying $\phi$ is relevant for $U(1)_{L_\mu-L_\tau}$ and $U(1)_{L_\mu+L_e-2L_\tau}$ benchmark models, since in those models the $V\to \nu\nu$ decays have a large branching ratio and thus $\tau \to \ell \phi, \phi \to 2V\to 4\nu$ (and $\tau \to \ell V, V\to 2\nu$) is an important channel. 

Belle-II performed a search for $\tau \to \lepton \phi_\text{inv}$~\cite{Belle-II:2022heu,Belle:2025bpu}.
In a dataset consisting of $736 \times 10^6$ tau pairs no statistically significant signal was found, and thus a $m_\phi$-dependent 90\% CL upper limit on the $\tau \to \lepton \phi_\text{inv}$ branching fraction was placed. For our benchmark $m_\phi \simeq 1$\,GeV the bound is~\cite{Belle-II:2022heu,Belle:2025bpu}
\begin{equation}
\label{eq:tau:ell:inv:exp}
    \mathcal{B}(\tau \to \lepton \phi_\text{inv}) < 0.3 (0.6) \times 10^{-3} \,\text{for } \lepton = \mu (e).
\end{equation}
Note that while the final state lepton is mono-energetic in the tau rest frame, this rest frame is typically not well reconstructed (the exception would be STCF working at tau threshold). 
This makes the searches for such invisible decays particularly challenging. 
Interpreting \cref{eq:tau:ell:inv:exp} as a bound on $\tau \to \ell 4\nu$ and $\ell 2\nu$, i.e., as a bound on $\mathcal{B}(\tau \to \lepton \phi)\cdot \mathcal{B}(\phi \to VV)\cdot \mathcal{B}^2(V\to \nu\bar{\nu}) + \mathcal{B}(\tau \to \lepton V)\cdot \mathcal{B}(V\to \nu\bar{\nu})$, 
the resulting constraint on $\Lambda$ for the Model~III benchmark is shown as the teal shaded region in \cref{fig:model3_Lambda} (right).

\paragraph{\bf Multibody $\boldsymbol{\tau \to \lepton+\mathrm{inv}}$ decays.}  The  $\tau \to \ell +3\cS$ decay in model V will appear as $\tau \to \ell+$inv decay in the limit when $\cS$ lifetime is large enough that it escapes the detector before decaying. For typical benchmark values considered for model V, this only occurs when $\cC_{e\cS}$ is very small, $\cC\lesssim 10^{-7}$, see \cref{eq:ctauLcS}. The multibody $\tau \to \ell +3\cS_\text{inv}$ decays face large SM backgrounds from the leptonic tau decays, $\tau \to \ell \nu \bar\nu$. Similarly to the two body $\tau \to \ell \phi_\text{inv}$ decays, one would need to rely on shape information to distinguish a possible signal from the SM background. Such searches have not yet been performed. However, even if one optimistically assumes that the resulting bound is no  worse than the one for the two body case, \cref{eq:tau:ell:inv:exp}, this leads to a relatively weak bound $\Lambda \gtrsim 0.5$\,TeV.

\paragraph{\bf The $\boldsymbol{\tau \to 3\lepton}$ decays.} 
The  $\tau \to \ell \ell' \ell'$ decays are classic new physics search channels. Most commonly, the bounds on these are interpreted in terms of constraints on heavy new physics, however, they also lead to constraints on models with light new physics. 
For instance, the current world leading limits on $\tau \to 3 \lepton$ decays were due to Belle II and are (all limits are at the 90\% CL):
\begin{subequations}
\begin{align}
&\mathcal{B}(\tau^- \to e^- e^+ e^-) < 2.7 \times 10^{-8},&\text{\cite{Hayasaka:2010np}}\\
&\mathcal{B}(\tau^- \to e^- \mu^+ \mu^-) < 2.7 \times 10^{-8},&\text{\cite{Hayasaka:2010np}}\\
&\mathcal{B}(\tau^- \to e^+ \mu^- \mu^-) < 1.7 \times 10^{-8},&\text{\cite{Hayasaka:2010np}}\\
&\mathcal{B}(\tau^- \to e^- \mu^- e^+) < 1.8 \times 10^{-8},&\text{\cite{Hayasaka:2010np}}\\
&\mathcal{B}(\tau^- \to e^- \mu^+ e^-) < 1.5 \times 10^{-8},&\text{\cite{Hayasaka:2010np}}\\
&\mathcal{B}(\tau^- \to \mu^- \mu^+ \mu^-) < 1.9 \times 10^{-8}.&\text{\cite{Belle-II:2024sce}}
\end{align}
\end{subequations}
In benchmark models I-IV, the $\tau^-\to \ell^- \ell^{\prime +} \ell^{\prime -}$ decays can be due to the $\tau\to \ell \phi$ or  $\tau\to \ell V$ two body decays, followed by $\phi \to 2\ell'$ and $V \to 2\ell'$, respectively. The $\ell$ are thus monoenergetic in the $\tau$ rest frame, while $2\ell'$ would form a resonant mass peak at $m_\phi$ or $m_V$ in their invariant mass distribution. Integrating over the three body phase space, the above searches 
place a limit on the combination
\begin{equation}
\mathcal{B}(\tau^- \to \lepton^- \phi)\cdot \mathcal{B}(\phi \to \ell'\ell')+\mathcal{B}(\tau^- \to \lepton^- V)\cdot \mathcal{B}(V \to \ell'\ell')~.
\end{equation}
The constraints are most relevant for larger values of flavor diagonal couplings of $\phi$ and $V$, so that a sizable fraction of them decays inside the detector. 
The resulting bounds on $\Lambda$ are shown as the orange shaded region in \cref{fig:model1_Lambda} (right) for Model~I, and as the orange ($\tau \to \mu\, e^+e^-$) and green ($\tau \to 3\mu$) shaded regions in \cref{fig:model3_Lambda} (right) for Model~III. 
Note that flavor violating decays, $\phi \to \ell \ell'$ or $V\to \ell \ell'$ are further suppressed, and thus the limits on  $\mathcal{B}(\tau^- \to \ell^- \mu^- e^+)$ 
and $\mathcal{B}(\tau^- \to \ell^- \mu^+ e^-)$ do not lead to meaningful constraints on the benchmark models I-IV.

\paragraph{\bf Multibody $\boldsymbol{\tau \to 3\lepton +}$inv decays.} In the SM, radiative corrections to the main semileptonic channel, $\tau \to \ell \nu_\tau \bar \nu_\ell$, 
give rise to the following multi-body decays
\begin{subequations}
\label{eq:tau:3ell:inv}
\begin{align}
    &\mathcal{B}(\tau^+ \to e^+ e^+ e^- \bar{\nu}_\tau \nu_e) \simeq 4.5\times 10^{-5}~,\\
    &\mathcal{B}(\tau^+ \to \mu^+ e^+ e^- \bar{\nu}_\tau \nu_\mu) \simeq 2.1 \times 10^{-5}~, \\
    &\mathcal{B}(\tau^+ \to \mu^+ \mu^+ \mu^- \bar{\nu}_\tau \nu_\mu) \simeq 1.3 \times 10^{-7}~.
\end{align}
\end{subequations}
The above SM branching ratios were estimated using
 \textsc{MadGraph}@{NLO}, in the approximation where the contribution from photon mixing with hadronic vector resonances can be ignored. 
Current best measurements of these channels are due to CLEO~\cite{CLEO:1995azm}
\begin{subequations}
\begin{align}
    \label{eq:tau:mu:2e:inv}
    &\mathcal{B}(\tau^- \to \mu^- e^- e^+ \bar{\nu}_\mu \nu_\tau) < 3.2 \times 10^{-5}, 
    \\
    \label{eq:tau:e:2e:inv}
    &\mathcal{B}(\tau^- \to e^- e^- e^+ \bar{\nu}_e \nu_\tau) = (2.8 \pm 1.5) \times 10^{-5},
\end{align}
\end{subequations}
in agreement with our approximate SM predictions. 

In order to avoid any dependence on the modeling of hadronic resonance contributions to the SM rates, we take a conservative approach of requiring that the new physics contributions do not violate the experimental bounds in \cref{eq:tau:3ell:inv}.  
In benchmark models II and III, when $\phi$ and $V$ decay inside the detector, \cref{eq:tau:mu:2e:inv} places constraints on the following product of branching ratios, 
\begin{equation}
2 \mathcal{B}(\tau \to \mu \phi)\cdot \mathcal{B}(\phi \to VV) \cdot \mathcal{B}(V \to \nu\bar{\nu})\cdot  \mathcal{B}(V \to 2e).
\end{equation}
while  \cref{eq:tau:e:2e:inv}, interpreted as a 90\%CL bound on the branching ratio, $\mathcal{B}(\tau^- \to e^- e^- e^+ \bar{\nu}_e \nu_\tau) < 4.7 \times 10^{-5}$, constrains
\begin{equation}
2 \mathcal{B}(\tau \to e \phi)\cdot \mathcal{B}(\phi \to VV) \cdot \mathcal{B}(V \to \nu\bar{\nu})\cdot  \mathcal{B}(V \to 2e).
\end{equation}
The impact of these constraints on the reach of Model~III is depicted in \cref{fig:model3_Lambda} (right) by the purple shaded region. 

If $V$ has a sufficiently long lifetime, the invisible signature in $\tau \to 3\ell+$inv can also arise from one $V$ escaping the detector while the other decays visibly inside it.
In Model~I, this requires $\varepsilon \lesssim 10^{-5}$, a region that is already strongly constrained by beam dump and fixed-target experiments searching for displaced dark photon decays~\cite{Ilten:2018crw,Baruch:2022esd}.
For Models~II and III, displaced $Z'$ decays require very small gauge couplings, $g' \lesssim 10^{-6}$, a regime where stellar cooling and cosmological constraints become increasingly relevant.
More generally, in all models I--IV, the same gauge coupling that controls the $V$ lifetime also governs the $\phi \to VV$ partial width.
Reducing it enough for displaced $V$ decays simultaneously suppresses $\Gamma(\phi \to VV)$, eventually driving $\phi$ to decay through subdominant channels such as $\phi \to \ell\ell'$, leaving $\tau \to 3\ell$ as the dominant cascade signature.

\paragraph{\bf Exotic Higgs decays.} The Yukawa-like interactions involving $\phi$, see \cref{eq:generalL:dim5,eq:generalL:dim6}, will induce exotic decays of the form 
$h\to \tau \mu \phi$. In the case of dimension 5 operator, \cref{eq:generalL:dim5}, the branching ratio is given by~\cite{Galon:2017qes}
\beq
\begin{split}
\frac{\mathcal{B}(h\to \tau \mu \phi)}{\mathcal{B}(h\to \tau\tau)}&=\frac{1}{6}\biggr(\frac{m_h c_{23}^{(5)}}{4\pi y_\tau \Lambda_5}\biggr)^2 
=0.66\cdot \biggr(\frac{0.5\,\text{TeV}}{\Lambda_5/c_{23}^{(5)}}\biggr)^2, 
\end{split}
\eeq
and similarly for $h\to \tau e \phi$, replacing $c_{23}^{(5)}\to c_{13}^{(5)}$. For simplicity only $c_{23}^{(5)}$ was assumed to be nonzero, while $c_{32}^{(5)}=0$. If this is not the case, one should replace $\big(c_{23}^{(5)}\big)^2\to \big(c_{23}^{(5)}\big)^2+\big(c_{32}^{(5)}\big)^2$ above. In the numerics we used the SM value of tau Yukawa, $y_\tau=1\cdot 10^{-2}$.
For a light invisible $\phi$, with a mass below 2\,GeV, the $h\to \tau \ell \phi$ decays  resemble  the $h\to \tau \ell$ decays in LHC searches \cite{Galon:2017qes}. Since tau decays always involve neutrinos, one cannot fully reconstruct the higgs mass, rather relying on approximate kinematical constraints. The additional missing energy in $h\to \tau \ell \phi$ broadens the distributions, but it is reasonable to expect that an ${\mathcal O}(1)$ fraction of events would pass the cuts (due to the use of BDT to optimize signal-to-background ratio in experimental searches, the exact efficiency for the signal is hard to estimate). 

Current limits on $h\to \ell \tau$ decays are
\begin{subequations}
\begin{align}
\mathcal{B}(h\to \tau \mu)&< 1.5 \cdot 10^{-3}~\text{\cite{CMS:2021rsq}},
\\
\mathcal{B}(h\to \tau e)&<2.0 \cdot 10^{-3}~\text{\cite{ATLAS:2023mvd}}.
\end{align}
\end{subequations}
Assuming as the best case scenario acceptance for  $h\to \tau \ell \phi$ signal is the same as for $h\to \tau \ell$ the above bounds would translate to $\Lambda_5> 2.6 (2.2)\,\text{TeV}$ for $h\to \tau \mu$ ($h\to\tau e$). Even in this optimistic case, the reach  is many orders of magnitude lower than the typical NP scale probed in tau decays, cf. \cref{eq:Br:tau:ell:phi:example}. If the $h\to \tau \ell \phi$ decays are induced by dimension 6 operators, \cref{eq:generalL:dim6}, the above discussion still applies, replacing everywhere $c_{ij}^{(5)}/\Lambda_5\to \sqrt{2} v' c_{ij}^{(6)}/\Lambda_6^2$. 

If $\phi $ decays within detector, $\phi\to 2\ell$, this would lead to exotic decays of the type $h\to \tau \ell 2\ell'$. While direct searches for such decays may lead to enhanced sensitivity~\cite{Evans:2019xer},  translating to a factor few larger probed values of NP scales $\Lambda_{5,6}$, they are still orders of magnitude below the parameter space probed by multilepton tau decays.

\paragraph{\bf Rare $\boldsymbol{\tau \to \lepton \gamma}$ decays.} At one loop two insertions of the effective operator \cref{eq:generalL:dim5} generate $\tau \to \lepton \gamma$ transition. For $\phi$ and $\ell'$ lepton running in the loop the resulting branching ratio is $\mathcal{B}(\tau \to \ell \gamma)\sim {\mathcal O}(10^{-30}) |c_{3\ell'} c_{\ell'\ell}|^2(10^{11} \,\text{GeV}/\Lambda_5)^4$  \cite{Galon:2017qes}, where as a numerical example we used the same value of $\Lambda_5$ as in \cref{eq:Br:tau:ell:phi:example} and the benchmark mass of $\phi$, $m_\phi=1\,$GeV, cf. \cref{eq:benchmark}. This is well below any current or future limits on $\tau \to \lepton \gamma$ \cite{BaBar:2009hkt}. 

\paragraph{\bf Lepton flavor universality.}
The measurement of lepton flavor universality ratio, 
\beq
R_{e/\mu}\equiv
\frac{\mathcal{B}(\tau \to e X_\text{inv})}{\mathcal{B}( \tau \to \mu X_\text{inv})},
\eeq
can be used to place limits on the difference between $c_{13}$ and $c_{23}$ Yukawa-like couplings in \eqref{eq:generalL}, if $\phi$ either decays invisibly, $\phi \to 2V\to 4\nu$ as in Models  II and III, or if $\phi$ is so weakly coupled that it avoids detection.
That is, the deviation is given by
\beq
\begin{split}
\label{eq:LFU:ratio}
\frac{R_{e/\mu}}{R_{e/\mu}\big|_\text{SM}}-1
=\frac{ \Delta \mathcal{B}(\tau \to \ell X_\text{inv})|_\text{NP}}{\mathcal{B}( \tau \to \ell X_\text{inv})}+\cdots 
\end{split},
\eeq
where $\Delta \mathcal{B}(\tau \to \ell X_\text{inv})=\mathcal{B}(\tau \to e X_\text{inv})- \mathcal{B}(\tau \to \mu X_\text{inv})$, and $\mathcal{B}( \tau \to \ell X_\text{inv})=\frac{1}{2}\big[\mathcal{B}( \tau \to e X_\text{inv})+\mathcal{B}( \tau \to e X_\text{inv})\big]$, while ellipses denote higher order terms in lepton flavor universality breaking (including the breaking present in the SM due to $m_\mu\ne m_e$). 
\Cref{eq:LFU:ratio} in principle sets bounds on the difference $|c_{13}|^2-|c_{23}|^2$. However, experimental limits are quite weak and place constraints on $\mathcal{B}(\tau \to \ell \phi_\text{inv})$ that are at the level of ${\mathcal O}(10^{-3})$, well above the interesting parameter range in \cref{eq:Br:tau:ell:phi:example}.

\paragraph{\bf Constraints from tau lifetime.}
The current world average of tau lifetime measurements, $\tau_\tau^\text{exp}=(290.3\pm0.5)\,$fs \cite{ParticleDataGroup:2024cfk} agrees well with the SM theory predictions, $\tau_\tau^\text{SM}=(288.6\pm2.3)\,$fs \cite{Erler:2002bu,Kretz:2025pfk}. 
The comparison of the two still allows for NP branching ratio, $\mathcal{B}(\tau\to X_\text{NP})<5\cdot 10^{-3}$, which is well above the branching ratios of relevance to exotic multi-lepton tau decays.

\paragraph{\bf Rare muon transitions.}
If the dark scalar $\phi$ has flavor violating couplings to both muons and electrons, $c_{13}, c_{23}\ne 0$, then $\mu \to e \gamma$ will be induced at one loop with $\phi$ and $\tau$ running in the loop. This process, however, requires two insertions of small flavor violating couplings, and is thus highly suppressed for typical values of the NP scale,  $\mathcal{B}(\mu \to e\gamma)\sim {\mathcal O}(10^{-25}) |c_{3e} c_{3 \mu}|^2(10^{11} \text{GeV}/\Lambda_5)^4$ \cite{Galon:2017qes} for benchmark $\phi$ mass, \cref{eq:benchmark}, well outside reach of any current or planned experiment. Similarly, the contributions to $(g-2)_\mu$ are highly suppressed, well below any future sensitivity. An exchange of $\phi$ coupling to dark photon (Model I), will also generate $(\bar \mu_L e_R) (\bar q q)$ four fermion operator at one loop. However, this operator is also highly suppressed, with Wilson coefficients proportional to  $\propto m_q g_D^2 \varepsilon^2/(16\pi^2 \Lambda_5 m_\phi^2)\sim (1/10^{11}\,\text{GeV})^2$, where in the numerics we used the benchmark values in \cref{eq:Br:tau:ell:phi:example,eq:benchmark,eq:benchmark:varepsilon}. This is well below the reach of current and any future planned muon conversion experiments \cite{Echenard:2026cdv}.

\paragraph{\bf Other collider constraints.} There are a number of constraints from beam-dumps and from collider searches for light scalar and bosons, constraining their flavor conserving couplings. Some of these we already mentioned in the introduction to this section, such as the bounds on flavor diagonal couplings of light vectors given in  \cite{Ilten:2018crw,Baruch:2022esd}. All our numerical benchmarks were chosen such that these bounds are satisfied.

The main sensitivity to flavor violating $\phi$\,--\,$\tau$\,--\,$\ell$ couplings are going to come from tau decays, as soon as the targeted searches for the multilepton tau decays are performed. 
The typical sensitivity to the NP scale suppressing the flavor violating couplings of the benchmark models is well above the direct collider reach, i.e., it is in the $10^3$ TeV regime for dimension 6 and is $\sim 10^{8}$\,TeV for dimension 5 operators. The exception is Model V, where searches for multilepton signatures will probe TeV scale suppressions of the $\tau$\,--\,$\ell$\,--\,$\cS^3$ couplings. UV completions of this model could thus in principle be searched for at LHC, though only if the full UV completion contains states that are not electroweak singlets and/or couple to quarks or gluons.  
 
A possible set of collider constraints is due to a quartic interaction between $\Phi$ and the SM Higgs, ${\cal L}\supset \lambda \Phi^2 |H|^2$, which induces $\phi$-Higgs mixing with the mixing angle, $\theta\sim \lambda v'/m_h$, after the Higgs and $\Phi$ obtain VEVs. This mixing then modifies the decay properties of $\phi$, the Higgs couplings to the SM fermions and gauge bosons, and induces exotic Higgs decays of the form $h\to 2\phi\to 4V\to 8\ell$. Demanding that the quartic coupling  $\lambda$ is small enough, $\lambda\lesssim 10^{-4}$, such that there is no significant tuning in the $\phi$ mass, however, implies that the mixing is very small, $\theta\lesssim 10^{-6}$, so that all these effects are unobservably small for current and planned colliders.

\section{Experimental prospects}
\label{sec:exp:prospects}
The experimental reach to new physics via searches for multi-lepton tau decays, such as $\tau \to 5\mu$, depends on how large these branching ratios are compared to other exotic tau decay channels, such as, e.g, $\tau \to 3\mu 2\nu$ or $\tau\to 3\mu2\pi$. These relative branching ratios are model dependent and one can increase  the sensitivity to particular model by covering more possible exotic decay channels listed in \cref{tab:allmodes}.

For each of the channels, the  experimental reach depends on the corresponding experimental efficiencies for that particular decay channels. Below we provide rough estimates of experimental sensitivities, by first estimating the 
expected number of taus 
 at relevant $pp$ and $e^+e^-$ colliders, and give approximate expected acceptance rates. These should be viewed only as order of magnitude estimates, motivating a more detailed study of experimental sensitivities for the most promising channels. 

\subsection{LHC searches} 
Multi-lepton tau decays can be efficiently searched for both at the general purpose detectors --- ATLAS and CMS --- as well as at LHCb.

The rare tau decay searches at LHCb benefit from its forward geometry ($2 < \eta < 5$), low-$p_T$ muon trigger threshold ($\sim 0.5$~GeV), and excellent vertex resolution.
Taus are produced copiously via heavy-flavor decays, predominantly $D_s \to \tau\nu$, with an effective cross section $\sigma_\tau^{\rm eff} \sim 30~\mu$b within the LHCb acceptance.
The LHCb Run~1+2 dataset ($\sim 9$~fb$^{-1}$) consists of approximately $10^{11}$ tau decays within LHCb's fiducial volume.
A preliminary analysis of $\tau \to 5\mu$ experimental reach at LHCb obtained a sensitivity of $\mathcal{B} \lesssim 4 \times 10^{-8}$~\cite{Overhoff:2020}.
At the HL-LHC with 300~fb$^{-1}$, the tau yield will increase to $\sim 2 \times 10^{12}$, which would mean that potentially branching ratios at the $10^{-10}$ level could be probed, assuming backgrounds can be controlled.

There are several important sources of backgrounds for such searches. An important consideration is the size of combinatorial background, which will increase at high luminosity where multiple $pp$ interactions per bunch crossing produce large backgrounds from random track combinations. This can be mitigated by requiring all five tracks to originate from a common displaced vertex consistent with the tau lifetime ($c\tau_\tau = 87~\mu$m). In the prompt regime we consider, the $\phi$ and $V$ decay lengths are negligible compared to this displacement, so all five tracks indeed emerge from a single point.
    
The $\tau \to 5\mu$ decay channel benefits from the clean muon signature, while channels involving pions (e.g., $\tau \to 3\mu 2\pi$) face significant backgrounds from prompt hadron production. Here, the resonant structure in the $m_{\pi\pi}$ mass distribution from $V \to \pi^+\pi^-$ decays, could provide discrimination, but dedicated studies are needed for sensitivity estimates. The most challenging are decays with missing energy; channels with neutrinos cannot be fully reconstructed at a hadron collider, making searches for $\tau \to 3\mu 2\nu$ and related signatures impractical at LHCb.

For ATLAS and CMS a significant limitation for such searches is the high-$p_T$ trigger threshold. That is, while both experiments have a larger instantaneous luminosity than LHCb, they also require muons with $p_T \gtrsim 20$\,--\,$26$~GeV at the hardware trigger level~\cite{ATLAS:2020gty,CMS:2020cmk}. 
For taus produced in heavy-flavor decays, the typical tau $p_T$ momentum is $\mathcal{O}(1)$~GeV, with only ${\mathcal O}(10^{-5})$ fraction of taus having $p_T$ above 25\,GeV \cite{Ema:2025bww}. The muons from $\tau\to 5\mu$ will thus be well below the nominal $p_T$ threshold. Alternatively, one could use taus with sufficiently high $p_T$ that are produced via $pp \to Z/\gamma^* \to \tau^+\tau^-$, but the cross section for this process ($\sim$~nb) is three orders of magnitude smaller than the heavy-flavor production. Furthermore, the boosted tau topology makes vertex reconstruction challenging.

 For dedicated searches, the trigger thresholds can be lowered significantly. For instance, for recent $\tau\to 3\mu$ search the CMS required two oppositely charged muons above $p_T>5(3)\,$GeV at L1 trigger, and at HLT that the third track is above $p_T>1.2\,$\,GeV \cite{CMS:2023iqy} (for previous searches at CMS, ATLAS and LHCb see \cite{CMS:2020kwy,ATLAS:2016jts,LHCb:2014kws}). An interesting question is whether additional trigger lines could be implemented for multimuon/multihadron tau decays or whether application of scouting \cite{CMS:2024zhe} or trigger level analyses \cite{ATLAS:2018qto} would be beneficial, in order to boost acceptance for such exotic signals.

\subsection{Electron-positron colliders}
\label{sec:e+e-:coll}
At $e^+e^-$ machines large tau sample are planned to be collected, $2\, (1.2)\times 10^{11}$ $\tau^+\tau^-$ pairs at FCC-ee \cite{Dam:2018rfz,Lusiani:2023LFVtau,Lusiani:2023FCCtau} (CEPC~\cite{CEPCStudyGroup:2018ghi,Ai:2024nmn}) running at the $Z$ pole,  $4.5\times 10^{10}$  $\tau^+\tau^-$ pairs at Belle II~\cite{Belle-II:2018jsg,Banerjee:2022xuw}, and $3.5\times 10^{10}$  $\tau^+\tau^-$ pairs at STCF~\cite{Ai:2024nmn}. Unlike the LHC searches the multi-lepton tau decays at $e^+e^-$ machines do not require a trigger. In the case of $Z$ pole measurements the challenge may be vertex reconstruction, due to the large $\tau$ boost in the lab frame ($\gamma_\tau\simeq 26$). Usually the large boost aids searches for exotic tau decays, since it makes it easier to approximately reconstruct the tau rest frame. However, in the case of fully visible final states, such at $\tau \to 5\mu$ this is not an issue, and rather the question is whether some of the muon tracks would merge in high-multiplicity events. It is expected that the large lepton multiplicity of the final state will lead to negligible backgrounds, similarly to the search for $\tau\to 3\mu$ at Belle II which resulted in current world best bound \cite{Belle-II:2024sce}. 
For the projected sensitivity curves in \cref{fig:model1_Lambda,fig:model3_Lambda}, we estimate branching ratio sensitivities at future facilities by rescaling the current best upper limits, assuming similar sensitivity per tau decay. Specifically, for channels with existing measurements, such as $\tau\to 3\mu$ and $\tau\to\mu e^+e^-$, the projected 90\% CL upper limits are assumed to scale as the inverse of the number of produced tau leptons, $\mathcal{B}_{\rm proj} = \mathcal{B}_{\rm current} \times N_\tau^{\rm current}/N_\tau^{\rm proj}$, as appropriate for signals with negligible backgrounds. For Belle II with 50 ab$^{-1}$ this gives, e.g., $\mathcal{B}(\tau\to 3\mu) \lesssim 1.6\times 10^{-10}$, rescaled from the current Belle II limit of $1.9\times 10^{-8}$ obtained with 424 fb$^{-1}$~\cite{Belle-II:2024sce}. For FCC-ee, we further rescale by the ratio of tau pair yields, $N_\tau^{\rm FCC\text{-}ee}/N_\tau^{\rm Belle,II} \approx 1.7\times 10^{11}/ 4.6 \times 10^{10} \approx 3.7$, giving $\mathcal{B}(\tau\to 3\mu) \lesssim 4.4\times 10^{-11}$. This procedure is conservative for FCC-ee, where the large tau boost ($\gamma_\tau\simeq 26$) is expected to improve reconstruction efficiencies for multi-body final states relative to $B$-factory energies.
Generically, we can expect  that $e^+e^-$ facilities can be sensitive to branching ratios for $\tau\to5\mu$ and other fully visible multi-lepton/multi-pion final states in the range ${\mathcal O}(10^{-9})$ to ${\mathcal O}(10^{-11})$. Note that this will translate to a reach of very high UV scales. For instance, $\mathcal{B}(\tau \to \mu \phi, \phi \to 4\mu)=10^{-11}$ probes a scale $\Lambda_5/c_{23}^{(5)}\simeq 3\cdot 10^{12}\,$GeV, see \cref{eq:Br:tau:ell:phi:example}. One could also search for partially invisible multilepton decays, such as $\tau\to 3\mu2\nu$. Such an analysis would be somewhat akin to the inelastic dark matter search at Belle II~\cite{Duerr:2019dmv,Liang:2022pul}.

At $e^+ e^-$ colliders one can also search for $\phi$ and $V$ via their flavor diagonal couplings. For instance, Belle  performed a search for $e^+e^- \to A' \phi \to A' A' A'$ \cite{Jaegle:2015fme}, while Belle II performed a search for $e^+e^- \to \mu \mu (X \to \mu \mu)$ \cite{Belle-II:2024wtd}. While these searches constrain the parameter space in our benchmark models, they are not sensitive to flavor off-diagonal couplings, and thus the high UV dynamics. 

\section{Conclusions}
\label{sec:conclusions}

Rare tau decays provide a sensitive probe of light new physics. 
In this work we explored dark sector models that produce multi-lepton signatures via cascade decays such as $\tau \to \ell\, \phi \to \ell\, VV \to \ell\, (XX)(YY)$, where $X, Y$ can be charged leptons, hadrons, or neutrinos. We have also explored the possibility of even higher multiplicity tau decays, such as $\tau \to 7\mu$ which can arise from $\tau \to \mu 3\cS, \cS\to 2\mu$ decay chains. 
Such multi-lepton final states are qualitatively distinct from the well-studied $\tau \to 3\ell$ decays and offer new experimental opportunities.

The initial step in the cascade, a higher-dimensional operator connecting two lepton flavors, $\ell_1, \ell_2$ and light particles $\phi,\,a,\, V,...$ can be called a leptonic FCNC portal to light states. Experimental studies of such portals are not as widespread as for the other lower-dimensional flavor-conserving portals \cite{Lanfranchi:2020crw}. Yet the wealth of new experimental signatures uncovered in this papers, as well as in the previous works addressing $(\ell_2\to \ell_1)$-initiated production of light states \cite{Calibbi:2020jvd,Hostert:2023gpk,Greljo:2025ljr}, and fairly large energy scales $\Lambda$ probed via such a portal, should serve as a strong motivation for further experimental searches of these portals. 

In this work, we focused on the prompt cascade regime, where the light new physics particles, $\phi, V$ or $\cS$ decay within the resolution of the tau vertex.
For benchmark couplings in the $\tau \to \ell\, \phi \to \ell\, VV$ decay chain examples,  $\varepsilon \sim 10^{-4}$ and $\alpha_d \sim \alpha$, the decay lengths are $c\tau_\phi \sim 100$\,fm and $c\tau_V \sim 10$\,--\,$30\,\mu$m, far smaller than typical vertex resolutions.
In this case, all five final-state particles thus emerge from a single displaced vertex, a signal that is well-suited for existing $\tau \to 3\ell$ analysis techniques with appropriate modifications.
The preliminary LHCb sensitivity for $\tau\to 5\mu$ is $\mathcal{B}(\tau \to 5\mu) \lesssim 4 \times 10^{-8}$ \cite{Overhoff:2020}, while at Belle II one can expect that the sensitivity is comparable to the current bound on $\tau \to 3\mu$, which is $\mathcal{B}(\tau \to 3\mu) < 1.9 \times 10^{-8}$ \cite{Belle-II:2024sce}. If the $\tau \to \mu\phi$ decay is due to a dimension 5 operator, this would then probe a NP scale $\Lambda_5\gtrsim 10^{11}$\,GeV, see \cref{eq:Br:tau:ell:phi:example}.

Exotic tau multilepton/multihadron decays remain almost completely unexplored experimentally. 
The benchmark models discussed in this manuscript predict eighteen distinct five-body channels, and eight seven-body decay channels, see \cref{tab:allmodes}. In contrast,  so far only $\tau \to 5\mu$ has a (preliminary) sensitivity study. 

Which of the multi-body channels dominate depends on the model. For instance, the hadronic channels $\tau \to 3\mu\,2\pi$ and $\tau \to \mu\,4\pi$ --- which can \textit{dominate} over $\tau \to 5\mu$ in Model~I above the two-pion threshold. That is, for $V$ a kinetically mixed vector  the $R$-ratio enhancement makes hadronic channels for $m_V > 2m_\pi$ the leading signatures.
Near the $\rho$ resonance, in particular, $\mathcal{B}(V \to \pi\pi) / \mathcal{B}(V \to \mu\mu) \sim 10$, so that $\mathcal{B}(\tau \to \mu\,4\pi) \gg \mathcal{B}(\tau \to 5\mu)$.

Whether or not the exotic decays include missing energy signatures can be an important guide to the symmetry structure of the new physics model. For instance, the signatures $\tau \to 3\mu + \slashed{E}$ and $\tau \to \mu + \slashed{E}$ arise from $V \to \nu\bar\nu$ decays in gauged lepton-number models, but are absent if $V$ is a kinetically mixed dark photon. Measuring ratios between different exotic decay branching fractions can thus discriminate between the models. That is, while  large $\tau \to 3\mu\,2\pi$ rate relative to $\tau \to 5\mu$ points to kinetic mixing, a comparable $\tau \to 5\mu$ and $\tau \to 3\mu\,2\nu$ rates indicate $L_\mu - L_\tau$; while $\tau \to 5e$ without $\tau \to 5\mu$ suggests a chiral model, at least among the benchmark models introduced here.
Even null results across multiple channels provide valuable discrimination.

While multi-lepton decays are exotic enough that the irreducible backgrounds are small, this may not be the case for accidental backgrounds in $pp$ collisions. The intermediate $V \to \ell^+\ell^-$ and $V \to \pi^+\pi^-$ decays produce peaks in the corresponding invariant mass distributions, which can aid in constructing best search strategies, distinguishing signal from backgrounds and directly measuring $m_V$ if observed.

While we focused on prompt decays, the benchmark models can also lead to displaced vertex signatures. For instance, for $\varepsilon \lesssim 10^{-5}$, the dark photon decay length exceeds $\mathcal{O}(\text{mm})$, producing resolvable displaced vertices. While more challenging, such displaced multi-track signatures are thus equally compelling.

Note that higher dimension FCNC operators may induce other decays in addition to tau ones. 
The details on this depend on the UV theory and flavor structure of the couplings. For instance in the dark photon model supplemented by higher dimension operators 
both couplings to quarks and leptons are allowed by $U(1)'$. That is, instead of the $\phi$ coupling to leptonic Yukawa operator as in  \cref{eq:generalL:dim5} one can equally well have couplings of $\phi$ to up or down Higgs Yukawa operators. This means that also signatures such at $B\to K4\ell $, $D\to \pi 4\ell,...$ could arise from similar cascade decays that we discussed in this paper (the possibility of such dark cascades from muon decays was already discussed in \cite{Greljo:2025ljr}).
Indeed, while there have been searches for $B_{(s)} \to 4\mu$ at LHCb~\cite{LHCb:2013zjg,LHCb:2021iwr} and $B^0 \to A'A' \to 4\ell$ at Belle~\cite{Belle:2020the}, there are no $B \to K A'A' \to K 4\ell$ searches to the best of our knowledge (see, e.g., \cite{Batell:2009jf}).
In other models, however, the dark cascades primarily occur only in tau decays. Examples of such model are the flavor protected scalar and $L_\mu-L_\tau$ with $Q_\Phi=+2$, in  which multilepton decays arise only from tau decays, while in $L_\mu+L_e-2L_\tau$, chiral $U(1)_\ell$ and $L_\mu-L_\tau$ with $Q_\Phi=+1$, both muon and tau dark cascade decays are allowed, but not dark cascade decays of heavy quarks. 

Many of these decays could be searched for at currently running experiments. Here, Belle~II is uniquely positioned to search the full spectrum of final states due to low backgrounds.
With 50\,ab$^{-1}$, sensitivities of $\mathcal{B} \sim 10^{-10}$ are achievable. The multi-lepton decay searches can also be performed at LHC experiments --- LHCb, CMS and ATLAS --- though dedicated or modified triggers may be required. 
Future $Z$-factories (FCC-ee, CEPC) with $\sim 10^{11}$ tau pairs can reach $\mathcal{B} \sim 10^{-11}$, which for FCNCs induced by dimension 5 operators would probe new physics scales in the $10^{12}$ GeV range.

In summary, exotic multi-lepton tau decays offer a largely unexplored window into light new physics.
The theoretical motivation is clear, the signatures are distinctive, and the searches are experimentally feasible. We encourage the experimental collaborations to pursue a comprehensive program across all predicted channels.

\begin{acknowledgements}
We thank Johannes Albrecht for communicating to us the preliminary LHCb sensitivity estimates.
This work was partially supported by the University of Iowa's Year 2 P3 Strategic Initiatives Program through funding received for the project entitled High Impact Hiring Initiative (HIHI): A Program to Strategically Recruit and Retain Talented Faculty. TM is supported in part by the Shelby Endowment for Distinguished Faculty at the University of Alabama and by Fermilab via Subcontract 725339. JZ acknowledges support in part by DOE grants DE-SC101977 and DE-SC0026301, and by NSF grants OAC-2103889, OAC-2411215, and OAC-2417682. AR acknowledges support from the Natural Sciences and Engineering Research Council of Canada (NSERC) and  Arthur B. McDonald Canadian Astroparticle Physics Research Institute. Research at Perimeter Institute is supported by the Government of Canada through the Department of Innovation, Science, and Economic Development, and by the Province of Ontario.
\end{acknowledgements}

\bibliography{rare_tau_multilepton}

\end{document}